\documentclass[sigconf]{acmart}

\AtBeginDocument{%
  }

\copyrightyear{2026}
\acmYear{2026}
\setcopyright{cc}
\setcctype{by-nc-nd}
\acmConference[CHI '26]{Proceedings of the 2026 CHI Conference on Human Factors in Computing Systems}{April 13--17, 2026}{Barcelona, Spain}
\acmBooktitle{Proceedings of the 2026 CHI Conference on Human Factors in Computing Systems (CHI '26), April 13--17, 2026, Barcelona, Spain}
\acmPrice{}
\acmDOI{10.1145/3772318.3790418}
\acmISBN{979-8-4007-2278-3/2026/04}

\acmSubmissionID{4015}

\usepackage{array}       
\usepackage{booktabs}    
\usepackage{cleveref}
\usepackage{xspace}
\usepackage{enumitem}

\definecolor{verylightgray}{RGB}{210, 210, 210}
\definecolor{veryverylightgray}{RGB}{245, 245, 245}
\definecolor{darkgray}{HTML}{4a4a4a}
\definecolor{xlcolor}{HTML}{008080}
\definecolor{charlescolor}{HTML}{ff6955}
\definecolor{fycolor}{HTML}{7765db}
\definecolor{mattcolor}{HTML}{4287f5}
\definecolor{annabelcolor}{HTML}{cc78ac}
\definecolor{NavyBlue}{HTML}{000000}
\usepackage{pifont}

\newif\ifnotes
\notesfalse 

\definecolor{charlescolor}{HTML}{ff6955}
\definecolor{mattcolor}{HTML}{4287f5}
\definecolor{r1color}{HTML}{A020F0}

\newcommand{\charles}[1]
{\ifnotes{\leavevmode\color{charlescolor}{(Charles: #1)}}\fi}

\definecolor{amethyst}{rgb}{0.6, 0.4, 0.8}

\usepackage{soul}

\newcommand{\etal}{\mbox{et\ al.}\xspace}

\newcommand{\eg}{\mbox{e.g.,}\xspace\@}

\newcommand{\llm}{{\textsc{llm}}\xspace}
\newcommand{\ai}{{\textsc{ai}}\xspace}

\usepackage{soul}
\usepackage[normalem]{ulem} 

\definecolor{inputULColor}{HTML}{aaaaaa}
\newcommand{\inputs}[1]{{\color{inputULColor}\uline{\textcolor{black}{\itshape{#1}}}}}

\definecolor{stageULColor}{HTML}{18a2ed}
\newcommand{\stages}[1]{{\color{stageULColor}\uline{\textcolor{black}{\itshape{#1}}}}}

\newcommand{\inlinequote}[1]{{\textit{#1}}}

\makeatletter
\newcommand\myparagraph[1]{%
  \@startsection{paragraph}{4}{\z@}%
  {.25em \@plus .25em \@minus.0em}%
  {-.5em}%
  {\bf}
  {#1}%
}
\makeatother

\newcommand{\ripplet}[0]{Ripplet\@\xspace}

\newcommand{\DOref}[2]{\hyperref[#1]{\textbf{#2}}}

\usepackage{graphicx}

\newif\ifshowedit
\showeditfalse

\definecolor{editred}{HTML}{ed0923}

\newcommand{\edit}[1]{%
  \ifshowedit
    {\color{editred}#1}%
  \else
    {#1}%
  \fi
}

\definecolor{editminorred}{HTML}{b5767d}

\definecolor{gray90}{HTML}{bbbbbb}
\newcommand{\remove}[1]{\ifshowedit{\leavevmode\color{gray90}\selectfont\sout{#1}}\fi}

\begin{document}


\title[Codesigning Ripplet, an LLM-Assisted Assessment Authoring System with Teachers]{Codesigning Ripplet: an LLM-Assisted Assessment Authoring System Grounded in a Conceptual Model of Teachers' Workflows}

\author{Yuan ``Charles'' Cui}
\email{yuancui2025@u.northwestern.edu}
\orcid{0000-0002-2681-6441}
\affiliation{%
  \institution{Northwestern University}
  \city{Evanston}
  \state{Illinois}
  \country{USA}
}

\author{Annabel Marie Goldman}
\email{annabelgoldman2025@u.northwestern.edu}
\orcid{0009-0004-5600-725X}
\affiliation{%
  \institution{Northwestern University}
  \city{Evanston}
  \state{Illinois}
  \country{USA}
}

\author{Jovy Zhou}
\email{jovyzhou@u.northwestern.edu}
\orcid{0009-0002-1040-3973}
\affiliation{%
  \institution{Northwestern University}
  \city{Evanston}
  \state{Illinois}
  \country{USA}
}

\author{Xiaolin Liu}
\email{xiaolinliu2026@u.northwestern.edu}
\orcid{0009-0009-5462-1711}
\affiliation{%
  \institution{Northwestern University}
  \city{Evanston}
  \state{Illinois}
  \country{USA}
}

\author{Clarissa M Shieh}
\email{clarissashieh2027@u.northwestern.edu}
\orcid{0009-0000-5657-5872}
\affiliation{%
  \institution{Northwestern University}
  \city{Evanston}
  \state{Illinois}
  \country{USA}
}

\author{Joshua Yao}
\email{joshuayao2026@u.northwestern.edu}
\orcid{0009-0008-3426-4975}
\affiliation{%
  \institution{Northwestern University}
  \city{Evanston}
  \state{Illinois}
  \country{USA}
}

\author{Mia Lillian Cheng}
\email{miacheng@gmail.com}
\orcid{0009-0008-2480-3330}
\affiliation{%
  \institution{Northwestern University}
  \city{Evanston}
  \state{Illinois}
  \country{USA}
}

\author{Matthew Kay}
\email{mjskay@northwestern.edu}
\orcid{0000-0001-9446-0419}
\affiliation{%
  \institution{Northwestern University}
  \city{Evanston}
  \state{Illinois}
  \country{USA}
}

\author{Fumeng Yang}
\email{fy@umd.edu}
\orcid{0000-0002-8401-2580}
\affiliation{%
  \institution{University of Maryland}
  \city{College Park}
  \state{Maryland}
  \country{USA}
}








\renewcommand{\shortauthors}{Cui et al.}

\begin{abstract}

Assessments are critical in education, but creating them can be difficult. To address this challenge in a grounded way, we partnered with 13 teachers in a seven-month codesign process. We developed a conceptual model that characterizes the iterative dual process where teachers develop assessments while simultaneously refining requirements. To enact this model in practice, we built Ripplet,\footnote{A demo video of the system is provided in supplemental materials.} a web-based tool with multilevel reusable interactions to support assessment authoring. The extended codesign revealed that \ripplet{} enabled teachers to create formative assessments they would not have otherwise made, shifted their practices from generation to curation, and helped them reflect more on assessment quality. In a user study with 15 additional teachers, compared to their current practices, teachers felt the results were more worth their effort and that assessment quality improved.

\end{abstract}


\begin{teaserfigure}
  \vspace*{-5pt}
  \includegraphics[width=\textwidth]{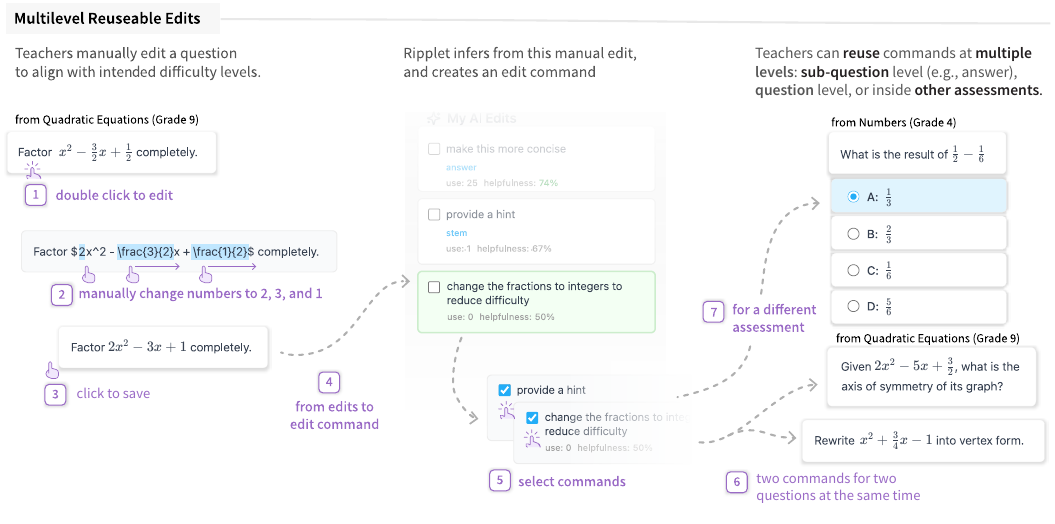}
  \vspace*{-26pt}
  \caption{An example feature of \ripplet{}'s multilevel reusable edits with \llm{}s: A teacher manually edits a question, and the system infers why this change was made and creates a reusable edit command, which can be reapplied to questions in other assessments.\looseness=-20}
  \Description{Figure 1: “This figure describes an example feature of Ripplet's multilevel reusable edits with LLMs. A teacher manually edits a question on quadratic equations (factor x^2-(3/2)x+1/2 completely) that contains fractions so that it only contains integers (2x^2-3x+1). After they save the manual changes, Ripplet uses an LLM to infer why this change was made and creates a reusable edit command called ``change the fractions to integers to reduce difficulty'', which the teacher can then select and reuse to apply to change other questions in any assessment, multiple questions at a time if they wish.”
}
  \label{fig:teaser}
\end{teaserfigure}
\DeclareRobustCommand{\duplicate}{%
  \begingroup\normalfont%
 \raisebox{-2pt}{\includegraphics[height=9pt]{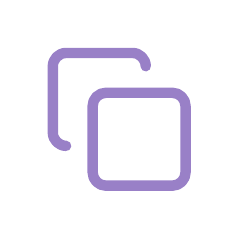}}%
  \endgroup%
}

\DeclareRobustCommand{\undo}{%
  \begingroup\normalfont%
 \raisebox{-2pt}{\includegraphics[height=9pt]{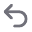}}%
  \endgroup%
}

\DeclareRobustCommand{\generateSimilar}{%
  \begingroup\normalfont%
 \raisebox{-2pt}{\includegraphics[height=9pt]{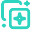}}%
  \endgroup%
}
\DeclareRobustCommand{\copyQuestion}{%
  \begingroup\normalfont%
 \raisebox{-1pt}{\includegraphics[height=9pt]{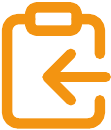}}%
  \endgroup%
}

\DeclareRobustCommand{\copyGrey}{%
  \begingroup\normalfont%
 \raisebox{-1pt}{\includegraphics[height=9pt]{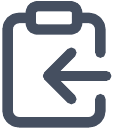}}%
  \endgroup%
}

\DeclareRobustCommand{\delete}{%
  \begingroup\normalfont%
 \raisebox{-1pt}{\includegraphics[height=9pt]{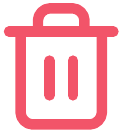}}%
  \endgroup%
}

\DeclareRobustCommand{\shuffle}{%
  \begingroup\normalfont%
 \raisebox{-1pt}{\includegraphics[height=9pt]{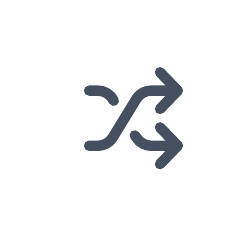}}%
  \endgroup%
}

\DeclareRobustCommand{\collapse}{%
  \begingroup\normalfont%
 \raisebox{-2pt}{\includegraphics[height=9pt]{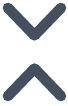}}%
  \endgroup%
}

\DeclareRobustCommand{\downloadQuestion}{%
  \begingroup\normalfont%
 \raisebox{-1pt}{\includegraphics[height=9pt]{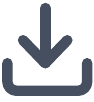}}%
  \endgroup%
}

\DeclareRobustCommand{\lightbulb}{%
  \begingroup\normalfont%
 \raisebox{-2pt}{\includegraphics[height=9pt]{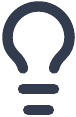}}%
  \endgroup%
}

\DeclareRobustCommand{\refinement}{%
  \begingroup\normalfont%
 \raisebox{-2pt}{\includegraphics[height=10pt]{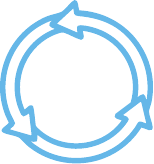}}%
  \endgroup%
}


\DeclareRobustCommand{\MIa}{%
  \begingroup\normalfont%
 \raisebox{-2pt}{\includegraphics[height=10pt]{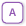}}%
  \endgroup%
}

\DeclareRobustCommand{\MIb}{%
  \begingroup\normalfont%
 \raisebox{-2pt}{\includegraphics[height=10pt]{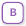}}%
  \endgroup%
}

\DeclareRobustCommand{\MIc}{%
  \begingroup\normalfont%
 \raisebox{-2pt}{\includegraphics[height=10pt]{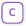}}%
  \endgroup%
}

\DeclareRobustCommand{\MId}{%
  \begingroup\normalfont%
 \raisebox{-2pt}{\includegraphics[height=10pt]{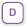}}%
  \endgroup%
}

\DeclareRobustCommand{\MIe}{%
  \begingroup\normalfont%
 \raisebox{-2pt}{\includegraphics[height=10pt]{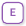}}%
  \endgroup%
}

\DeclareRobustCommand{\MIf}{%
  \begingroup\normalfont%
 \raisebox{-2pt}{\includegraphics[height=10pt]{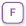}}%
  \endgroup%
}

\DeclareRobustCommand{\MIg}{%
  \begingroup\normalfont%
 \raisebox{-2pt}{\includegraphics[height=10pt]{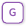}}%
  \endgroup%
}

\DeclareRobustCommand{\MIh}{%
  \begingroup\normalfont%
 \raisebox{-2pt}{\includegraphics[height=10pt]{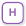}}%
  \endgroup%
}

\DeclareRobustCommand{\MIi}{%
  \begingroup\normalfont%
 \raisebox{-2pt}{\includegraphics[height=10pt]{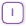}}%
  \endgroup%
}

\DeclareRobustCommand{\MIj}{%
  \begingroup\normalfont%
 \raisebox{-2pt}{\includegraphics[height=10pt]{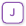}}%
  \endgroup%
}

\DeclareRobustCommand{\MIk}{%
  \begingroup\normalfont%
 \raisebox{-2pt}{\includegraphics[height=10pt]{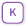}}%
  \endgroup%
}

\DeclareRobustCommand{\MIl}{%
  \begingroup\normalfont%
 \raisebox{-2pt}{\includegraphics[height=10pt]{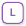}}%
  \endgroup%
}

\DeclareRobustCommand{\import}{%
  \begingroup\normalfont%
 \raisebox{-1pt}{\includegraphics[height=9pt]{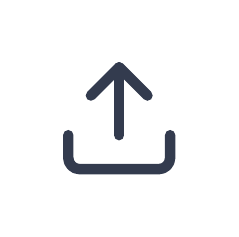}}%
  \endgroup%
}

\DeclareRobustCommand{\genimg}{%
  \begingroup\normalfont%
 \raisebox{-1pt}{\includegraphics[height=9pt]{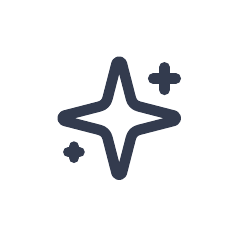}}%
  \endgroup%
}

\DeclareRobustCommand{\gentopic}{%
  \begingroup\normalfont%
 \raisebox{-2pt}{\includegraphics[height=9pt]{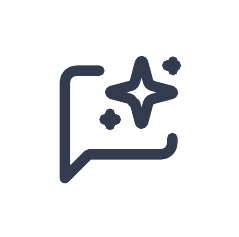}}%
  \endgroup%
}

\DeclareRobustCommand{\gencg}{%
  \begingroup\normalfont%
 \raisebox{-1pt}{\includegraphics[height=9pt]{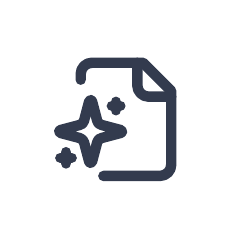}}%
  \endgroup%
}

\DeclareRobustCommand{\manual}{%
  \begingroup\normalfont%
 \raisebox{-2pt}{\includegraphics[height=9pt]{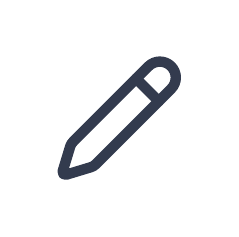}}%
  \endgroup%
}


\maketitle
\clubpenalty=0      
\widowpenalty=0     
\displaywidowpenalty=0
\raggedbottom

\section{Introduction}

Classroom assessments are critical in education~\cite{brown2005assessment, Remesal2011primary, barnes2017us-e34}. 
Formative assessments, such as daily homework and quizzes, help students learn and teachers monitor student progress~\cite{Remesal2011primary,taras2009summartive,black2009formative,brown2005assessment,brown2004teachers}. 
Summative assessments, such as final exams, measure student achievement and provide certification for broader audiences (\eg parents, school administrators)~\cite{taras2009summartive,Remesal2011primary}. 
Together, effective assessments facilitate learning and teaching \cite{black1998inside,gardner2006assessment, brown2005assessment, black2009formative}
and hold all shareholders accountable in the education ecosystem \cite{barnes2017us-e34,brown2005assessment, brown2006}.



However, developing assessments is
a time- and knowledge-intensive process~\cite{mcmillan2003assessment,wang2022towards,kurdi2020,stiggins1988revitalizing}.
Teachers need to create new questions, adapt existing materials, ensure formatting consistency, and align assessments with learning objectives~\cite{tierney2012fairness,ghaicha2016theoretical,deLuca2016}.
Yet some teachers receive insufficient training or support for assessment development \cite{stiggins1988revitalizing,mertler2009}.
Overworked and under-supplied teachers can produce low-quality assessments, creating inaccurate measures of students' abilities, which can potentially exacerbate disparities between high- and low-resource schools 
~\cite{kurdi2020, stiggins1988revitalizing, tarrant2006flaws, downing2005consequences}.

Prior work has studied how to reduce teachers' burden of developing assessments by helping them generate individual questions. Automatic question generation (\textsc{aqg}) aims to fully automate question creation with minimal teacher intervention. These methods include template-based generation~\cite{willert2024template} and more recently \llm-based approaches~\cite{cui2025vila,maity2024gpt4,maity2024prompt}. While \llm-based approaches have shown promise in generating questions across diverse contexts, they often struggle to produce high-quality questions that
can measure higher cognitive abilities ~\cite{maity2024gpt4,cui2025vila,gao2018,cheng2021guiding, fergus2023}. Recent work has studied how to involve teachers in \textsc{llm}-based approaches to produce high-quality questions, demonstrating the potential of human--\llm collaboration \cite{kang2025tutor,ReadingQuizMaker}. However, these works mostly focus on generating individual questions and do not provide sufficient insight into teachers' current assessment authoring workflows. As teachers need to create assessments (rather than individual questions), addressing this gap is essential to build tools that can integrate into their practice.
Our goal is to understand teachers' assessment authoring workflows and build a tool to support them through human-\textsc{llm} collaboration. To do so, we partnered with 13 teachers from eight U.S. high schools in a three-phase codesign process over seven months. In Phase I, we conducted formative interviews to understand teachers' assessment authoring workflows and created a conceptual model of teachers' assessment authoring practices (\Cref{subsec:conceptual_model}). The model describes assessment authoring as an iterative dual process in which teachers (1) create, reuse, and adapt content at multiple levels (\eg question level, assessment level) while also (2) refining their requirements about the content of the assessment. We derived from the model a set of design objectives for helping teachers adapt content, review and structure assessments, ensure quality, and integrate requirements. 

In Phase II (\Cref{sec:phase2}), we translated the conceptual model and design objectives into a system to explore how such a model could be enacted in practice. Through two rounds of design iteration, we developed \ripplet{}, a web-based tool for authoring assessments that embodies the process described in our model and allows us to observe how these ideas play out in real assessment authoring scenarios. Through a multilevel reusable interaction paradigm with \llm{s}, Ripplet supports creating questions from diverse inputs (\eg a list of topics, PDF of a curriculum), reusing and adapting content at multiple levels (sub-question, question, assessment, cross-assessment), restructuring assessments, and inferring, tracking, and reapplying teachers' edits. \Cref{fig:teaser} shows one of the key features of this paradigm---multilevel reusable edits---which support iterative authoring: teachers can edit an entire question or just a part of it either manually or by prompting an \llm. The system infers reusable commands from teachers’ edits. These commands can be reapplied to multiple questions or specific parts of questions in any assessment, supporting the multilevel nature of assessment authoring.\looseness=-1

Over seven months, we collected feedback from codesign partners' independent adoption of \ripplet{} and observed how their usage and perceptions evolved (\Cref{sec:phase3}). Seven teachers used \ripplet{} in their classrooms and administered assessments to their students. We found that Ripplet was integrated into teachers' diverse workflows and helped them in their iterative dual process of authoring assessment and refining requirements. Some reported that \ripplet{} helped them create formative assessments they would not have had time to make. Over time, teachers' usage shifted from generation to curation with a growing sense of ownership, and they became more reflective on assessment design and quality. We also conducted a user study with 15 additional secondary school teachers, comparing their current practices (control) to using \ripplet{} (\Cref{sec:user_study}). Compared to control, teachers found using ripplet to be a better experience that resulted in higher-quality assessments: enjoyment ($\mu=+2.70$ with 95\% CI of $[0.86,4.54]$), exploration ($\mu=+2.43$ $[0.56,4.31]$), results worth effort ($\mu=+1.93$ $[0.33,3.53]$), and assessment quality ($\mu=+1.11$ $[0.10,2.11]$) all improved on a $10$-point scale.\looseness=-1



Through this work, we contribute:
\begin{itemize}[leftmargin=19pt,itemsep=0pt, topsep=0pt,rightmargin=10pt]
\item A conceptual model of teachers' assessment authoring workflows, capturing their iterative dual process with multiple types of input and stages of action;
\item A system instantiation, Ripplet, that realizes the conceptual model through multilevel reusable interactions; 
\item Insight from extended codesign, showing that Ripplet supports the workflows in our model and that its multilevel reusable interactions eased assessment authoring while encouraging teachers to reflect on assessment design; 

\item Findings from a controlled study, showing how \ripplet{} improves authoring experience and assessment quality.
\end{itemize}

\vspace{5pt}
We also discuss the limitations of our codesign partner sample and how to build sustainable and reciprocal codesign practices with teachers.



\section{Related Work}
\subsection{Assessment Development}
Developing assessments is a time-intensive process~\cite{mcmillan2003assessment,wang2022towards,kurdi2020,stiggins1988revitalizing}, 
 consuming up to half of teachers' professional time~\cite{stiggins1986classroom, stiggins2001littletraining}.
They need to align assessments with learning objectives, devise intellectually demanding tasks, and ensure instructional relevance~\cite{tierney2012fairness,ghaicha2016theoretical,deLuca2016}.  
However, some teachers receive insufficient training or support for assessment development \cite{stiggins1988revitalizing,mertler2009}, potentially leading to low-quality assessments and consequently uneven chances among students to demonstrate their knowledge~\cite {kurdi2020, stiggins1988revitalizing, tarrant2006flaws, downing2005consequences}.

 
 
Researchers in education, psychology, and cognitive science have proposed measurement theories and models to guide assessment development, such as constructing detailed test blueprints \cite{psychTestingBook,blueprint-authenticass} and estimating item parameters (e.g., difficulty) \cite{national-testing,psychometrics-item-analysis,haladyna2013developing}.
Some argue that models of cognition and learning can inform educational assessment development, making assessments more effective in measuring student understanding \cite{nation-academy-ass}. However, translating such research into practice is not easy \cite{mcmillan2003assessment,nation-academy-ass}. In reality, teachers rarely adopt such principles when creating assessments, as these theories often lack relevance to their day-to-day instruction \cite{mcmillan2003assessment,AssessmentInTheAgeOfAI}. Consequently, existing work in test development theory provides insufficient practical guidance for how teachers currently develop assessments. In our work, we first build a conceptual model of teachers' assessment authoring practices through formative interviews (\Cref{sec:phase1}) and then ground the design of our system in the conceptual model.


\subsection{Automatic Question Generation }
Automatic question generation (\textsc{aqg}) methods aim to ease the burden of creating questions ~\cite{coombs2018, mulla2023, kurdi2020}, gaining momentum with the recent advances in generative \ai~\cite{mulla2023, kurdi2020}. 
These methods range from template-based generation to neural and transformer models~\cite{pan2020semantic,tuan2020q,xie2020exploring,uto2023difficulty,raina2022multiple,tuan2020,wang2018} and from student-authoring~\cite{moore2022} to web-scraping techniques~\cite{gong-etal-2022-khanq,chen2018learningq}.
Recent \llm-based methods dramatically lower the technical barrier to entry~\cite{stefan2023physics,hang2024mcq,maity2024chain,cheng2024tree, moore2022,fergus2023,elkins2023}, and can take inputs like Bloom's Taxonomy and textbooks ~\cite{maity2024gpt4,cui2025vila,maity2024prompt,maity2024exploring}  to generate questions of varying difficulty levels~\cite{chen2018learningq,cheng2021guiding} 
 in multiple choice, free response, and fill-in-the-blank formats~\cite{mulla2023}. 
These methods typically achieve reasonable performance on quality metrics~\cite{ling2024question,bulathwela2023,maity2024prompt,maity2024chain, raina2022multiple,wang2018,cheng2021guiding} and human ratings (\eg  answerability)~\cite{steuer2021,moore2022,maity2024prompt,maity2024chain,wang2018,cheng2021guiding,elkins2023}.

However, there exists a gap between these automated methods and teachers' practices. Most automated solutions are not designed for teachers: some are end-to-end pipelines that require technical expertise (\eg coding) to execute, while others use generic interfaces that lack targeted support for assessment authoring. In addition, effective assessments must closely align with individual teachers' curricula, institutional guidelines, and the unique needs of student cohorts~\cite{pastore2019,deLuca2010,elkins2023}\textemdash{}requirements that cannot be met by sifting through thousands of generic questions. Assessment authoring is also a deeply personal practice that embodies teachers' pedagogical philosophy, beliefs, and years of accumulated knowledge about their students~\cite{coombs2018,xu2016,looney2018identity,adie01022013}. 
Automated methods without careful design to involve human oversight do not fit real-world classrooms.

\begin{figure*}[h]
    \includegraphics[width=\textwidth]{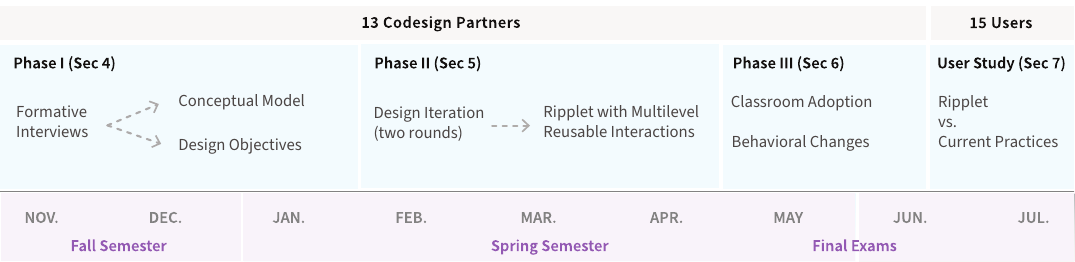}
    \caption{Overview of the three-phase codesign process and the controlled user study.}
    \label{fig:codesign-overview}
    \Description{Figure 2: “This figure provides an overview of the three-phase codesign process (13 codesign partners) and the controlled user study (15 users) from Nov to JUl. From Nov to Jan is Phase I, which overlaps with the fall semester. In Phase I (Sec 4), we conducted formative interviews to create a conceptual model and design objectives. From Jan to May is Phase II (Sec 5), which spans the spring semester, and we developed Ripplet with multilevel reusable interactions through two rounds of design iterations. From end of May and start of Jun is Phase III (Sec 6), where teachers adopted Ripplet in their classrooms and we observed their behavioral changes over time. From mid June to start of July is the user study with 15 additional teachers, where we compared Ripplet with their current assessment authoring practices.”}
\end{figure*}

\subsection{Human--AI Collaboration for Question Generation and Multilevel Iterative Problems}
Prior work has explored ways to incorporate human oversight into \llm-based question generation ~\cite{ReadingQuizMaker,kang2025tutor,cheng2024tree,fan2024lessonplanner,promptTheHive}. TutorCraftEase allows teachers to generate questions with \llm{}s and accept, reject, and manually edit \llm outputs. This enables teachers to produce quality questions more efficiently~\cite{kang2025tutor}. Similarly, ReadingQuizMaker helps college instructors make questions for reading quizzes, providing control for when to request \ai assistance; the authors found that instructors preferred this collaborative approach over an \ai-only process ~\cite{ReadingQuizMaker}.
While these systems show promise in mitigating the issues of fully automated approaches, they operate largely at the level of individual questions and provide limited support for reusing and adapting questions. Assessment authoring is inherently \textbf{multilevel}: it requires not only generating individual questions but also structuring them into a holistic assessment, tweaking specific parts of a question to fit students' needs, balancing difficulty and topic coverage, and aligning the collection to curricular goals. As we find in formative interviews (\Cref{sec:phase1}), assessment authoring is also \textbf{\textbf{iterative}}: teachers move between creating and adapting questions, searching for relevant materials, reviewing and restructuring the content, and refining their requirements for an assessment. 

Researchers in \textsc{hci} have studied multilevel problems. For example, CoLadder, a code generation system, allows users to construct a tree of prompt blocks, each block representing a different level of abstraction. This multilevel structure allows programmers to break down high-level goals into subtasks and procedural details, making the \ai-generated code easier to navigate and edit across those levels~\cite{CoLadder}. In addition, DirectGPT tackles multilevel editing by enabling users to apply changes to a specific part of an object, resulting in less time spent and more preferred outcomes than standard chat interfaces ~\cite{directgpt}. \textsc{hci} researchers  have also designed techniques to address iterative problems with \ai. PromptHive, a prompt authoring interface that supports rapid iteration on prompt variations, helps teachers create hints for educational questions by allowing prompts to be shared and reused~\cite{promptTheHive}. Similarly, ABScribe, an \llm-based system for writing, supports prompt reuse by turning edits into ``\ai modifiers'' that can be reapplied and edited~\cite{abscribe}. Recent work by Huang \etal and Zhang \etal demonstrates how \llm{}s can help solve iterative problems by translating low-level \textsc{ui} interactions into relevant reusable macros and higher-level goals ~\cite{autoMacro, SummAct}. Another solution to iterative problems is co-adaptive \ai systems where users and systems refine their behavior in tandem \cite{Yiming2025-AdaptiveLLMCollaboration, Jianhui2025TwinCo, Feng2025Canvill}. Twin-Co, for example, improves image generation through multi-turn editing, updating output as users specify their intent \cite{Jianhui2025TwinCo}. We draw on these interaction techniques to design a system with multilevel reusable interactions to support assessment authoring. \looseness=-1

\section{Codesign Overview}

To design a system that integrates into teachers' practices, we need to ground our design process in their lived experiences and obtain feedback over an extended period. Therefore, we engaged in a \textbf{participatory design} process with a diverse group of teachers (\Cref{fig:codesign-overview}). This process consisted of \textbf{three phases} over \textbf{seven months} in a typical U.S. academic year followed by our codesign partners’ schools, which comprises a fall semester (August/September – December) and a spring semester (January/February – May/June): in Phase I (\Cref{sec:phase1}), we conducted formative interviews to understand teachers' assessment authoring workflows and challenges. From there, we created a conceptual model of teachers' authoring practices and defined a set of design objectives for our system. In Phase II (\Cref{sec:phase2}), we conducted two rounds of design iteration where teachers used our prototypes and provided feedback that guided refinements toward the final system. We completed the final version before the next phase. In Phase III (\Cref{sec:phase3}), we invited teachers to independently use \ripplet{} to create an assessment and administer it to their students at the end of the school year. Afterwards, we conducted follow-up interviews where they reflected on their experience, their students’ reactions, and the overall codesign process. 
After codesign, we recruited 15 additional teachers for a controlled user study
(\Cref{sec:user_study}).


\subsection{Participants}\label{sec:codesign_participants} We began recruiting by emailing schools that have previous partnerships with our institutions. 
We then used snowball sampling by asking early recruits to recommend colleagues and friends.
We arrived at 13 codesign partners, including 6 female and 7 male teachers from 8 schools (6 public and 2 private) across 3 U.S. states, with teaching experience from 2 to 28 years; \Cref{tab:codesign_partners} describes their characteristics. Everyone was 18 years or older, spoke fluent English, and was actively teaching at the time of the study. 
The entire codesign was approved by our Institutional Review Board (\textsc{irb}),\footnote{We withhold the \textsc{irb} approval number for anonymized submission.} and each participant received electronic gift cards as compensation (Phase I: 40 \textsc{usd}; Phase II: 60 \textsc{usd} per session; Phase III: 90 \textsc{usd}).\footnote{We increased compensation in later rounds to encourage continued participation. 
} \looseness=-1


\aboverulesep=0ex
\belowrulesep=0ex
\setlength\arrayrulewidth{.2pt}
\bgroup
\begin{table*}[]
\centering
\caption[]{\textbf{Our Codesign Partners' Demographics and Participation.} We have a diverse cohort of partners from a wide range of years of teaching experience, school types, and subjects. 
}
\small
\vspace*{-8pt}
\def\arraystretch{1}%
\label{tab:codesign_partners}
\Description{Table 1: “Table titled “Our Codesign Partners’ Demographics and Participation.” Columns (left to right): Id, Sex, Years of teaching, State, School type, Subjects, and participation in four activities (Formative Interview, Design Iteration \#1, Design Iteration \#2, Independent Use), where “Yes” indicates participation. Row P1: Sex F; Years 22; State OH; School Private; Subjects Algebra, Pre-Calculus, Calculus; Formative Interview Yes; Design Iter. \#1 Yes; Design Iter. \#2 Yes; Independent Use Yes. Row P2: Sex M; Years 26; State OH; School Private; Subjects History, Geography, Economics; Formative Interview Yes; Design Iter. \#1 Yes; Design Iter. \#2 Yes; Independent Use Yes. Row P3: Sex M; Years 28; State OH; School Private; Subjects Biology, Ecology; Formative Interview Yes; Design Iter. \#1 Yes; Design Iter. \#2 Yes; Independent Use Yes. Row P4: Sex M; Years 27; State IL; School Public; Subjects Chemistry; Formative Interview Yes; Design Iter. \#1 Yes; Design Iter. \#2 Yes; Independent Use No. Row P5: Sex M; Years 23; State IL; School Public; Subjects Physics, Biology; Formative Interview Yes; Design Iter. \#1 No; Design Iter. \#2 Yes; Independent Use No. Row P6: Sex F; Years 28; State IL; School Public; Subjects Biology; Formative Interview Yes; Design Iter. \#1 Yes; Design Iter. \#2 Yes; Independent Use No. Row P7: Sex M; Years 22; State IL; School Public; Subjects Biology; Formative Interview Yes; Design Iter. \#1 No; Design Iter. \#2 Yes; Independent Use Yes. Row P8: Sex F; Years 9; State IL; School Public; Subjects Biology, Health Careers; Formative Interview Yes; Design Iter. \#1 Yes; Design Iter. \#2 Yes; Independent Use Yes. Row P9: Sex M; Years 6; State OH; School Public; Subjects Pre-Calculus, Calculus; Formative Interview Yes; Design Iter. \#1 Yes; Design Iter. \#2 Yes; Independent Use Yes. Row P10: Sex F; Years 2; State OH; School Public; Subjects Geometry, Algebra; Formative Interview Yes; Design Iter. \#1 No; Design Iter. \#2 No; Independent Use No. Row P11: Sex F; Years 6; State OH; School Public; Subjects Algebra, Pre-Calculus; Formative Interview Yes; Design Iter. \#1 No; Design Iter. \#2 Yes; Independent Use Yes. Row P12: Sex M; Years 11; State OH; School Public; Subjects Algebra, Computer Science; Formative Interview Yes; Design Iter. \#1 No; Design Iter. \#2 Yes; Independent Use No. Row P13: Sex M; Years 26; State NY; School Private; Subjects History; Formative Interview Yes; Design Iter. \#1 Yes; Design Iter. \#2 Yes; Independent Use No. Summary: 13 partners (6 female, 7 male); Years of teaching range 2–28 (median around low-20s); States: IL (6), OH (6), NY (1); School types: Public (9), Private (4). Participation counts — Formative Interview: 13 Yes, 0 No; Design Iteration \#1: 8 Yes, 5 No; Design Iteration \#2: 11 Yes, 2 No; Independent Use: 6 Yes, 7 No.”}

\begin{tabular}{>{\arraybackslash}m{0.04\textwidth}
                m{0.04\textwidth}
                m{0.04\textwidth}
                m{0.04\textwidth}
                m{0.06\textwidth}
                m{0.25\textwidth}
                m{0.09\textwidth}
                m{0.06\textwidth}
                m{0.06\textwidth}
                m{0.06\textwidth}}
\midrule\\[-6pt]
\textbf{Id} &
\textbf{Sex} &
\textbf{Years} &
\textbf{State} &
\textbf{School} &
\textbf{Subjects} &
\textbf{Formative Interview} &
\textbf{Design Iter. \#1} &
\textbf{Design Iter. \#2} &
\textbf{Indep. Use} \\ [2pt]

P1  & F & 22 & OH & Private & Algebra, Pre-Calculus, Calculus & \ding{51} & \ding{51} & \ding{51} & \ding{51} \\
P2  & M & 26 & OH & Private & History, Geography, Economics  & \ding{51} & \ding{51} & \ding{51} & \ding{51} \\
P3  & M & 28 & OH & Private & Biology, Ecology               & \ding{51} & \ding{51} & \ding{51} & \ding{51} \\
P4  & M & 27 & IL & Public  & Chemistry                      & \ding{51} & \ding{51} & \ding{51} &           \\
P5  & M & 23 & IL & Public  & Physics, Biology               & \ding{51} &           & \ding{51} &           \\
P6  & F & 28 & IL & Public  & Biology                        & \ding{51} & \ding{51} & \ding{51} &           \\
P7  & M & 22 & IL & Public  & Biology                        & \ding{51} &           & \ding{51} & \ding{51} \\
P8  & F & 9  & IL & Public  & Biology, Health Careers        & \ding{51} & \ding{51} & \ding{51} & \ding{51} \\
P9  & M & 6  & OH & Public  & Pre-Calculus, Calculus         & \ding{51} & \ding{51} & \ding{51} & \ding{51} \\
P10 & F & 2  & OH & Public  & Geometry, Algebra              & \ding{51} &           &           &           \\
P11 & F & 6  & OH & Public  & Algebra, Pre-Calculus          & \ding{51} &           & \ding{51} &           \\
P12 & M & 11 & OH & Public  & Algebra, Computer Science      & \ding{51} &           & \ding{51} & \ding{51} \\
P13 & M & 26 & NY & Private & History                        & \ding{51} & \ding{51} & \ding{51} &           \\

\midrule
\end{tabular}
\vspace*{-10pt}
\end{table*}
\egroup

\subsection{General Procedure}
All codesign sessions were conducted one-on-one on Zoom. 
Before the first session, we obtained consent to record both the audio and video 
and provided them with the option to turn off their video. 
To facilitate conversations during these sessions, we used seed questions\footnote{The seed questions for each phase are provided in supplemental materials.} that probed their practices while leaving space for them to raise questions. 

\myparagraph{Phase I: Formative Interviews.} Each partner participated in an hour-long semi-structured interview to share their assessment authoring practices and challenges. We transcribed all interviews and conducted a thematic analysis~\cite{thematic-analysis} on all transcripts. Two authors open-coded all the transcripts. The first author then created an affinity diagram \cite{affinity-diagram} based on the initial coding to extract themes. A third author then reviewed and commented on themes, and these three authors iteratively refined, split, and merged the themes through several rounds of discussion to reach a consensus ~\cite{hci-qual-reliability}.\footnote{As the initial coding is not the product but the process to generate themes, we do not compute measures such as inter-rater reliability \cite{hci-qual-reliability}.} The result is our conceptual model in \Cref{subsec:conceptual_model} and design objectives in \Cref{subsec:design_objectives}.

\myparagraph{Phase II: Design Iteration.} After developing the initial prototype of \ripplet{}, we invited our partners to participate in two rounds of design iteration sessions. Each session began with a brief tutorial---covering the entire system in the first round and new functionalities in the second---and was followed by hands-on use of \ripplet{} while thinking aloud. 
At the end of each session, teachers participated in a semi-structured interview reflecting on their experiences, likes, confusions, and suggestions for refinement. 
During these sessions, we took written notes on what participants did while using the system, the issues they verbalized while thinking aloud, and their responses during the interview. We also recorded the video and audio of these sessions so we could later reference them to supplement our written notes.
Over three months, we continually iterated on the design of \ripplet{} between each codesign session using our observations of teachers' usage and their direct feedback. The result of Phase II is the final version of \ripplet{} (\Cref{sec:phase2}).


\myparagraph{Phase III: Classroom Adoption.} Towards the end of the academic year, we invited our partners to use \ripplet{} independently to create an assessment they would later administer to their students as a practice exam or as the official final exam. 
After they completed this task, we conducted follow-up interviews in which they reflected on their independent use, their students’ feedback, and their overall experiences with \ripplet{} across the codesign process. 
A thematic analysis (same as Phase I) of the interviews gave us insight into how teachers used the system in authentic classroom settings (\Cref{sec:independent_classroom}), and also allowed us to capture longitudinal observations on how our codesign partners interacted with the system over an extended period of time (\Cref{subsec:behavioral}).


\section{Codesign Phase I: Conceptual Model and Design Objectives}\label{sec:phase1}

We conducted formative interviews with our codesign partners. From there, we developed a conceptual model of their authoring practices and derived design objectives for our target system. 


\subsection{Conceptual Model for Assessment Authoring}\label{subsec:conceptual_model}
\begin{figure*}[t]
\includegraphics[width=\textwidth]{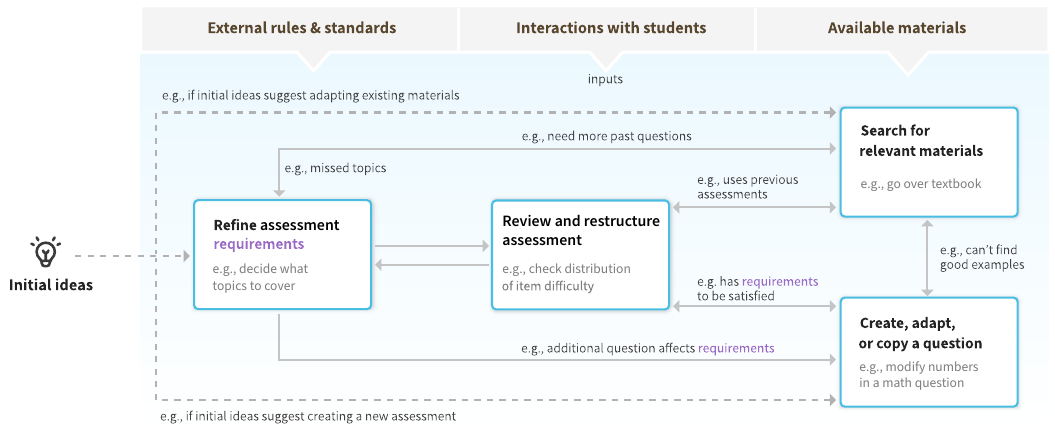}
\vspace{-15pt}
\caption{\textbf{Our conceptual model for teachers' processes of assessment authoring}.
In this model, a teacher holds some initial ideas about the assessment requirements. They then start and move between the \stages{stages} to simultaneously develop the assessment and refine assessment requirements, while considering various \inputs{inputs}.}\label{fig:conceptual_model}
\Description{Figure 3: “This figure presents a conceptual model of teachers’ processes of assessment development. At the left, an icon labeled “Initial ideas” enters the model. The background is divided into three columns across the top: (1) External rules \& standards, (2) Interactions with students, (3) Available materials. Inside a dashed box (representing the process), there are four main rectangular nodes: 1. “Refine assessment requirements” (purple subtitle). Example text: “e.g., decide what topics to cover.” 2. “Review and adjust assessment structure.” Example text: “e.g., check distribution of item difficulty.” 3. “Search for relevant materials.” Example text: “e.g., go over textbook.” 4. “Create, adapt, or copy a question.” Example text: “e.g., modify numbers in a math question.” Arrows connect these boxes in multiple directions, showing iteration: - From “Refine assessment requirements” to “Review and adjust assessment structure,” with example text above the arrow: “e.g., missed topics.” - A bidirectional arrow between “Refine assessment requirements” and “Review and adjust assessment structure.” - From “Review and adjust assessment structure” to “Search for relevant materials,” with arrow label: “e.g., need more past questions.” - From “Search for relevant materials” to “Create, adapt, or copy a question,” with arrow label: “e.g., can’t find good examples.” - From “Create, adapt, or copy a question” to “Review and adjust assessment structure,” with arrow label: “e.g., has requirements to be satisfied.” - From “Create, adapt, or copy a question” back to “Refine assessment requirements,” with arrow label: “e.g., additional question affects requirements.” - From “Review and adjust assessment structure” to “Refine assessment requirements,” with arrow label: “e.g., missed topics.” - From “Review and adjust assessment structure” to “Search for relevant materials,” with arrow label: “e.g., uses previous assessments.” Additional context labels around the edges: - Top left arrow from “Initial ideas” into the dashed box says: “e.g., if initial ideas suggest adapting existing materials.” - Bottom dashed arrow says: “e.g., if initial ideas suggest creating a new assessment.” Overall, the figure shows teachers iteratively moving between refining requirements, adjusting structure, searching for materials, and creating or adapting questions, while considering external rules, student interactions, and available materials.”}
\end{figure*}

Our conceptual model shows that assessment authoring is an iterative dual process (\Cref{fig:conceptual_model}) in which
teachers (1) \textbf{develop the assessment} while they simultaneously (2) refine its \textbf{requirements}. In this process, teachers juggle three categories of \inputs{inputs} while moving between four major \stages{stages} of action.



\subsubsection{Inputs} We identify three categories of \inputs{inputs} to assessment authoring: (1) external rules and standards, (2) interactions with current students, and (3) available materials. 
\myparagraph{\inputs{External Rules \& Standards}}
Teachers do not have complete autonomy in designing assessments; they must adhere to rules and standards from curricula, schools, districts, or states. In states like Illinois, the Next Generation Science Standards~(\textsc{ngss}) define what students should know and how they demonstrate understanding in K--12 science education~\cite{ngss}. For example, P8 created biology tests emphasizing data analysis to align with \textsc{ngss}'s focus on cross-cutting skills. Standardized, high-stakes exams also exert a strong influence: Advanced Placement (\textsc{ap}) is a program that offers university-level curricula and exams to high school students in the U.S. and Canada~\cite{ap}. 
Teachers often mimic the \textsc{ap} exam formats in their tests <P1, 3, 4, 9>. 
As P4 noted, ``\inlinequote{we do try to mirror the \textsc{ap} exam ... that exam will have some multiple choice and some free response.}'' Beyond standardized content, logistical constraints such as limited exam time and quick grading requirements further restrict assessment development: ``\inlinequote{I've gotta turn grades around within 24 hours, so I can't do a bunch of open-ended questions ... I have to write a multiple-choice test}'' <P5>.

\myparagraph{\inputs{Teacher's Interactions with Current Students}}
Teaching is a continuous interaction in which teachers observe students' progress through classroom activities and formative assessments (e.g., quizzes), then adapt assessments to suit students' specific strengths, struggles, backgrounds, and expectations. As P11 noted, ``\inlinequote{I'm spending a lot of my job watching how my kids do the problem ... when I make the tests, I know I wanna ask the type of questions that they've been tripped up on ... I wanna see that they're getting past that.}'' 

\myparagraph{\inputs{Available Materials}}
When creating assessments, teachers typically draw from various resources, including self-curated question banks, colleague-shared materials, official curricula (e.g., released past \textsc{ap} exams), and online sources (e.g., teacher forums). While these materials often prove useful, their quantity, quality, and relevance vary, influencing how much teachers need to adapt or create new content. As P5 noted, ``\inlinequote{I'm not sure where they're sourcing their questions ... their diagrams are really outdated ... I'll have like an idea of what question I want to ask, and then I'll go to Google Images ... then I will just re-tailor the question to fit that diagram.}''

\subsubsection{Actions} 
 Teachers rarely define all requirements---topics to cover, difficulty levels, question formats, test length, and alignment with \inputs{students' experiences} <P1, 9, 12, 13>---at the outset. Instead, they engage in an \textbf{iterative dual} process: they (1) develop the assessment while they simultaneously (2) refine its requirements. Some begin with a vague sense of test topics, refining those topics as they explore \inputs{available materials} <P1, 2, 12>. Others draw on semi-developed requirements from past courses and focus on creating or adapting questions, making occasional adjustments to their requirements for current \inputs{students} <P8, 9>. 
 We split this dual process into four stages: one for refining requirements, and three for developing the assessment\looseness=-1. 

\myparagraph{\stages{Refine Assessment Requirements}} 
Teachers begin the authoring process with a set of initial ideas about their requirements, which evolve as they develop the assessment in the dual process outlined above. Updating the requirements may prompt teachers to \stages{search} from \inputs{available resources} or \stages{create} new questions, and to \stages{review and} \stages{restructure} the assessment for any gaps. For instance, after reviewing past exam performance, P8 ``\inlinequote{either change\emph{[}s\emph{]} the questions or \emph{[}gives\emph{]} students a little bit more support and scaffolding}'' to improve a new test. Teachers continue refining requirements throughout the process until the final assessment meets their goals.

\myparagraph{\stages{Search for Relevant Materials}} 
If teachers' initial ideas point them to start by adapting \inputs{available materials}, they may begin by searching for relevant content, such as course topics, past tests, homework, textbook problem sets, curriculum-prescribed question banks, or online resources: ``\inlinequote{I can look through \emph{[}old textbooks\emph{]} to quickly glance and see ... that looks like it matches what I'm trying to assess from my students}'' <P1>. From here, teachers may transition to \stages{create, adapt, or copy}---adapting or copying if they find relevant materials or creating a new question if they do not.
In this stage, teachers may also spot gaps that prompt them to \stages{review and restructure} \stages{the assessment} or \stages{refine requirements}: ``\inlinequote{I kind of look through \emph{[}my notes and grade book\emph{]} and say, am I forgetting anything? Is there anything that we taught that we spent time on that I'd like to test these students on?}'' <P12>. Teachers often return to the search stage repeatedly throughout the authoring process to seek relevant materials.

\myparagraph{\stages{Create, Adapt, or Copy a Question}} 
Teachers may also begin by creating new questions if their initial ideas suggest so.
This is common for first-time teachers who may not have sufficient available materials to adapt <P2, 12> or those who prefer novelty and worry about test leakage\footnote{Test leakage is the release or sharing of test materials that could compromise the fairness and integrity of the test; e.g., if past years' students share tests.}~<P4, 13>. For instance, P13 tried to create new questions that differ from \mbox{\inputs{students' expectations}:} ``\inlinequote{I try to make ... a question that presents itself a little bit differently than exactly what everybody thinks is coming.}'' 
Teachers also adapt or copy past questions. This is often less laborious, especially for experienced teachers. If they find a helpful question by \stages{searching} from their \inputs{available materials}, they can copy it to the work-in-progress assessment or adapt it to fit the new one; for example, P7 said: ``\inlinequote{I've never given the same test ever ... at the very least I take a previous question and change some variables to it so they can do it in a ... different context.}'' As teachers add questions to the assessment, they regularly \stages{review and restructure} it to decide what to do next.

\myparagraph{\stages{Review \& Restructure Assessment}} 
If teachers find a past test from \stages{searching available materials} or have a work-in-progress assessment, they may review and restructure it based on their current requirements, such as coverage of topics, adherence to \inputs{external} \inputs{standards}, or \inputs{characteristics of current students}: ``\inlinequote{I tweak \emph{[}old tests\emph{]} every year ... I make sure that I cover the standards and the things that I've stressed within that semester}'' <P9>. 
If teachers find the current assessment does not satisfy their requirements, they may \stages{search} for materials or \stages{create, adapt, or copy} questions to address the gaps: ``\inlinequote{I pulled up the last couple years \emph{[}of final exams\emph{]} ... we got further this year ... So the final exam had to account for a test on cell rests. So we also looked at some cell rest questions that we were gonna add to the final to make it a little beefier there}'' <P7>. Teachers revisit this stage until all requirements are met, which concludes the authoring process.


\subsection{Design Objectives}\label{subsec:design_objectives}
The conceptual model illustrates assessment authoring as a complex process. 
Our interviews revealed challenges in efficiently using these inputs, completing each stage, and moving fluidly between stages. 
For example, teachers often struggled to efficiently adapt materials across contexts, or to revise assessments without duplicating effort. 
Building on our conceptual model, we distill a set of design objectives for an assessment authoring system.  

\begin{figure*}
    \centering
        \includegraphics[width=\linewidth]{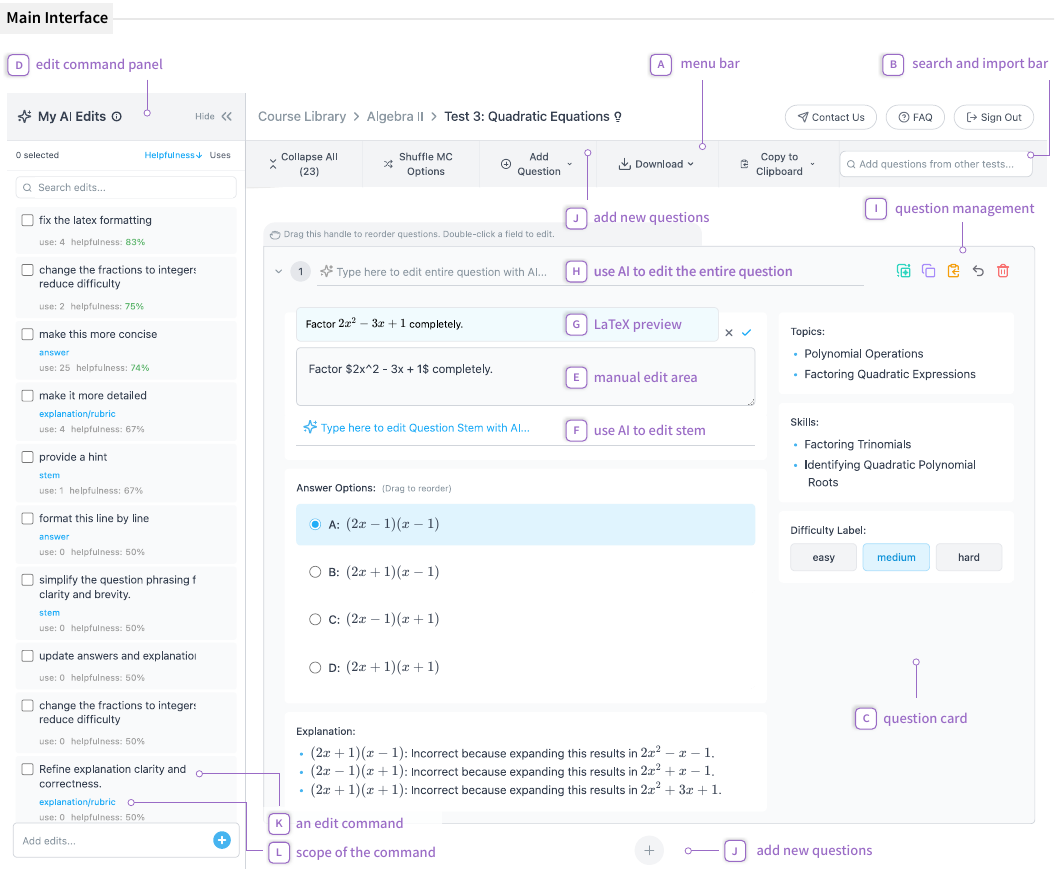}
    \captionof{figure}{
    The main interface (after double-clicking the question stem) of \ripplet{} consists of four major components: \MIa\ menu bar, \MIb\ search and import bar, \MIc\ question cards, and \MId\ edit command panel. The menu bar \MIa\ allows users to add questions and navigate, restructure, and export assessments. The search and import bar \MIb\ allows users to look for questions from other assessments in \ripplet{} and add to the current assessment. Below \MIa\ and \MIb\ lives the questions in the assessment. \ripplet{} supports two formats of questions: multiple-choice and free-response. Each question is displayed as a question card \MIc\, which has six sections: question stem, answer options (multiple-choice) / answer (free-response), explanation (multiple-choice) / suggested rubric (free-response), topics, skills, and difficulty label. Users can edit a part of a question manually \MIe\ or with \ai \MIf. A preview window \MIg\ is shown for content that contains \LaTeX. User can also edit the entire question with \ai \MIh. The five icons \MIi\ allow users to manage individual question and create additional ones: using \ai to generate similar questions, duplicating, copying to clipboard, undoing an edit, and deletion. Users can also add questions \MIj. On the left side is the edit command panel \MId. This panel stores the edit commands \MIk\ a user has made to the questions. Each command has a scope tag \MIl\ that specifies the part of the question it will change once applied.}
    \label{fig:main_interface}
    \Description{Figure 4: “This figure shows the main interface of Ripplet (after double-clicking the question stem). The caption identifies four major components: (A) the menu bar, (B) the search and import bar, (C) the question cards, and (D) the edit command panel. On the top, the menu bar (A) contains buttons: Collapse All (23), Shuffle MC Options, Add Question, Download, and Copy to Clipboard. On the right side of this bar are Contact Us, FAQ, and Sign Out. The search and import bar (B) allows adding questions from other tests. Below, the center area displays question cards (C). A sample question is shown: “Factor 2x^2 - 3x + 1 completely.” The LaTeX preview area shows “Factor $2x^2 - 3x - 1$ completely.” The manual edit area lets the user type text directly, while the AI edit stem option lets AI generate or revise text. The answer options include four multiple-choice items: (1) (2x - 1)(x - 1), (2) (2x + 1)(x - 1), (3) (2x - 1)(x + 1), (4) (2x + 1)(x + 1). Below the options is an explanation: - (2x - 1)(x - 1): Incorrect because expanding results in 2x^2 - 3x + 1. - (2x + 1)(x - 1): Incorrect because expanding results in 2x^2 - x - 1. - (2x - 1)(x + 1): Incorrect because expanding results in 2x^2 + x - 1. - (2x + 1)(x + 1): Incorrect because expanding results in 2x^2 + 3x + 1. Each question card also shows a right-hand panel with metadata: Topics (Polynomial Operations, Factoring Quadratic Expressions), Skills (Factoring Trinomials, Identifying Quadratic Polynomial Roots), and a Difficulty Label with buttons easy, medium, hard. On the left-hand side of the interface is the edit command panel (D) labeled “My AI Edits.” It lists a set of suggested edits with helpfulness scores, such as: “fix the latex formatting” (83\%), “change the fractions to integer reduce difficulty” (79\%), “make this more concise” (74\%), “make it more detailed explanation/rubric” (72\%), “provide a hint” (+59\%), “format this line by line answer” (50\%), “simplify the question phrasing if clarity and brevity” (+50\%), “update answers and explanations” (50\%), “change the fractions to integer reduce difficulty” (50\%), and “Provide explanation and clarity correctness” (50\%). The figure annotations label additional elements: E = Manual edit area, F = Use AI to edit stem, G = LaTeX preview, H = Use AI to edit entire question, J = Add new questions, K = an edit command, L = Scope of the command, I = Question management icons (generate similar, duplicate, copy to clipboard, undo, delete). The caption text explains: the menu bar (A) allows users to navigate, restructure, create and export assessments. The search and import bar (B) allows users to look for questions from other assessments and add them. Ripplet supports multiple-choice and free-response formats. Each question card (C) has six sections: question stem, answer options, explanation, suggested rubric, topics, skills, and difficulty. Users can edit questions manually (E) or with AI (F, H). The LaTeX preview (G) renders math. The icons (L) manage duplication, undo, copy, and delete. Users can also insert new questions at any location (I, J). The edit command panel (D) lists all AI edits with scope tags (K) that specify what part of the question will change.”}
\end{figure*}

\myparagraph{DO1: Enable Efficient Reuse and Adaptation of Assessment Content.}\phantomsection\label{do1}
Teachers rarely create assessments entirely from scratch; instead, they \stages{reuse and adapt content} from different types of \inputs{inputs} depending on their requirements—for example, using curriculum standards or targeting specific topics. As they move through the process, their requirements evolve, creating the need for efficient ways to \stages{create and adapt questions} to match updated requirements. Some teachers may need to fill unmet requirements by creating questions on a missing topic, while others want to create similar questions to an existing one to emphasize a concept. Yet this process is often fragmented and inefficient. For example, P12 explained the repetitive burden of this scattered process: ``\inlinequote{I’ll pull up the last couple years [of exams] … and then try to piece them together, but it’s a lot of cutting and pasting.}'' Teachers often have to \stages{search} through multiple documents, reformat content by hand, or recreate questions when small adjustments would have sufficed. A system should not only make it easy to surface content from existing sources but also help teachers adapt that material.

\myparagraph{DO2: Facilitate Assessment Evaluation and Restructuring.} \phantomsection\label{do2}
Teachers emphasized that it is critical to \stages{review and} \stages{restructure} an assessment to satisfy their requirements, yet this is cumbersome in existing tools. Teachers working in Google Docs or Word reported spending a large amount of time copying and pasting questions to reorder them. Some also expressed the difficulty of randomizing options for correct answers in multiple-choice questions on a test. A system should help teachers evaluate and restructure assessments fluidly without tedious copy-pasting.

\myparagraph{DO3: Ensure Assessment Quality.} \phantomsection\label{do3}
Teachers emphasized that maintaining the quality of assessments is essential but challenging. Many noted problems with existing materials: for example, P4 described how question banks often contain incorrect and low-quality questions, forcing them to solve problems himself to identify usable questions. Those who experimented with \ai tools such as ChatGPT voiced similar concerns, pointing out that hallucinations and unclear phrasing made it necessary to carefully verify correctness <P3–5, 9–12>. A system should help teachers \stages{review} and validate questions to ensure assessment quality.

\myparagraph{DO4: Integrate Requirements into Authoring Workflow.} \phantomsection\label{do4} 
Teachers described requirements as central to assessment authoring, yet in their current practices these requirements are rarely embedded directly into the authoring workflow. Instead, they often exist as separate artifacts—such as checklists, outlines, or mental notes—that must be manually cross-referenced during assessment authoring. Some requirements are relatively explicit and straightforward to track, such as ensuring balanced coverage of topics across a test. For instance, P5 described printing out their syllabus and marking off topics one by one to verify coverage. Other requirements are more implicit and harder to articulate directly, such as ensuring that questions are written at an appropriate reading level for students <P8>. Meeting these implicit requirements often demands significant additional effort and subjective judgment. This separation makes it easy for requirements to be overlooked and adds extra overhead to an already-demanding process. A system should provide lightweight support to ensure that an assessment satisfies a teacher's requirements.






\subsection{Challenges and Opportunities for AI Support}
While these design objectives are not specific to \ai, recent advances in \textsc{ai}—particularly large language models (\textsc{llm}s)—offer opportunities to help teachers achieve them. For example, prior work has shown that \textsc{llm}s can generate assessment questions given text or image input, which can help teachers reuse assessment content (\DOref{do1}{DO1}). Our partners also experimented with tools like ChatGPT to draft or edit questions. However, teachers also encountered clear limitations with existing \ai solutions. For instance, many found chat-based interfaces cumbersome for making targeted edits (\DOref{do1}{DO1}), often scrolling back and forth to locate specific questions or switching between multiple tools (e.g., Google Docs, ChatGPT) to adapt, label, or remove content <P1, 9, 10, 12>. The same limitation also creates a barrier for checking hallucinations and ensuring the quality of assessments (\DOref{do3}{DO3}). These challenges demonstrate that \ai alone is not enough to achieve these design objectives. Instead, it requires carefully designing human--\ai interactions so that \ai functions are embedded in workflows that reflect real authoring practices and provide support for teachers to evaluate and refine outputs to ensure assessment quality. 

\section{Codesign Phase II: System Description of \ripplet{}}\label{sec:phase2}

To realize the conceptual model and design objectives, we developed \ripplet{}, a system that leverages \textsc{llm}s to support the iterative dual process of assessment authoring.
In Phase II, we engaged in two rounds of design iteration with our partners and used their feedback to refine prototypes and arrive at the final system.
As we refined the design, we converged upon a multilevel reusable interaction paradigm, capable of generating questions from diverse inputs, adapting content at multiple levels, restructuring artifacts, and inferring, tracking, and reapplying edits. This paradigm supports question creation (\Cref{subsec:qgen}), basic question-level (\Cref{subsec:q-level}) and assessment-level (\Cref{sec:ass-level-ops}) operations, multilevel reusable edits (\Cref{subsec:multi-reuse-edits}), and cross-assessment operations (\Cref{subsec:crossassess}). We describe Ripplet's functions and their connections to the design objectives, highlight how Ripplet supports the conceptual model by marking the relevant \inputs{inputs} and \stages{stages}, and report major refinements (\refinement)\footnote{We do not report minor refinements such as adjustments to colors or sizes. We provide interface snapshots from past versions in supplemental materials.} from design iteration. The main interface consists of four major components (\Cref{fig:main_interface}): \MIa\ Menu Bar, \MIb\ Search and Import Bar, \MIc\ Question Cards, and \MId\ Edit Command Panel.

\subsection{Question Creation}\label{subsec:qgen}


To help teachers reuse content (\DOref{do1}{DO1}), \ripplet{} offers five methods for creating and importing questions. Per our conceptual model, teachers often use different inputs---such as \inputs{curriculum standards}, \inputs{past exams}, or \inputs{course topics}. There are four options to add questions  (\Cref{fig:main_interface} \MIj): teachers can (1) write questions by hand, or use \textsc{llm}s to (2) generate questions from topics and (3) curriculum guides, or (4) import past assessments. Teachers can also (5) \generateSimilar\ generate variations of an existing question on that question card. \refinement\ This method was added after the first round of design iteration, when P1, 4, 6, 9 expressed the need to create variations of existing, known-high-quality questions.

\subsection{Question-Level Operations}\label{subsec:q-level}
\ripplet{} offers a range of question-level operations that allow teachers to \stages{edit and manage individual questions}. 

\myparagraph{Manual Edit.} 
Double-clicking on a part (e.g., question stem) of a question triggers an inline editor for users to make direct, manual changes (\Cref{fig:main_interface}\,\MIe)
from the main interface, allowing teachers to adapt questions quickly (\DOref{do1}{DO1}).

\myparagraph{Question Management and Version Control.} 
\ripplet{} provides direct actions for managing each question (\Cref{fig:main_interface}\,\MIi).
\duplicate\ Duplication and \undo\ undoing changes help teachers efficiently reuse and \stages{adapt questions} across different versions (\DOref{do1}{DO1}), while \delete\ deleting allows them to remove weak or irrelevant questions to maintain  quality (\DOref{do3}{DO3}). \copyQuestion\ Copying a question to the clipboard allows teachers to export preferred versions and repurpose them in new contexts. These operations enable teachers to explore question variations while maintaining control and supporting iterative refinement.\looseness=-1

\subsection{Assessment-Level Operations}\label{sec:ass-level-ops}
Beyond individual questions, \ripplet{} supports operations that apply to an entire assessment (\raisebox{-2pt}{\includegraphics[height=9pt]{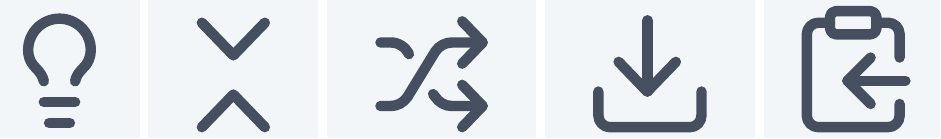}}), helping teachers efficiently adapt, organize, and distribute their assessments (\Cref{fig:main_interface}\,\MIa).

\myparagraph{Restructuring.}
To help \stages{restructure assessments} (\DOref{do2}{DO2}), \ripplet{} allows teachers to reorder questions via drag and drop. \shuffle\ Shuffling MC options randomizes the order of options within each multiple-choice question, helping reduce answer pattern bias and quickly create alternative versions of an assessment.

\myparagraph{Overview and Navigation.} Hovering over \lightbulb\ next to the assessment name displays the composition of questions by format and difficulty, helping teachers \stages{understand what requirements need to be} \stages{satisfied} (\DOref{do4}{DO4}). Teachers can \collapse\ collapse or expand all question cards to move through large assessments during evaluation and restructuring (\DOref{do2}{DO2}).

\myparagraph{Exporting.}
Teachers can \downloadQuestion\ download a PDF with or without answer keys and explanations/suggested rubrics. Teachers can also \copyGrey~copy and paste the assessment into other software (e.g., Word or Google Docs). Expressions, equations, and special symbols are preserved as inline images to ensure that they display as expected when pasted in other software. 

\subsection{Multilevel Reusable Edits with LLMs}\label{subsec:multi-reuse-edits} 
In addition to the basic operations at the question and assessment levels, \ripplet{} provides multilevel reusable edits which allow teachers to \stages{adapt questions} and reuse edits at the sub-question, question, and assessment level.

\begin{figure}
  \includegraphics[width=188pt]{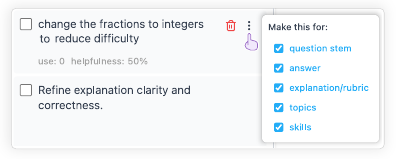}
\vspace*{-12pt}
  \captionof{figure}{Each command can be tagged with specific parts of a question to ensure that only those parts are modified when the command is applied.}
  \label{fig:tags}
  \Description{Figure 7: “This figure shows an example of edit commands in Ripplet. The panel lists two commands: (1) “change the fractions to integers to reduce difficulty,” marked with zero uses and a helpfulness score of 50\%. (2) “Refine explanation clarity and correctness.” Next to the first command, a dropdown menu titled “Make this for:” is open. It displays checkboxes for specific parts of a question that the command can target: question stem, answer, explanation/rubric, topics, and skills. All five options are checked in this example. The caption explains that each command can be tagged with specific parts of a question so that only those parts are modified when the command is applied.”}
\end{figure}

\begin{figure*}[t]
\includegraphics[width=\textwidth]{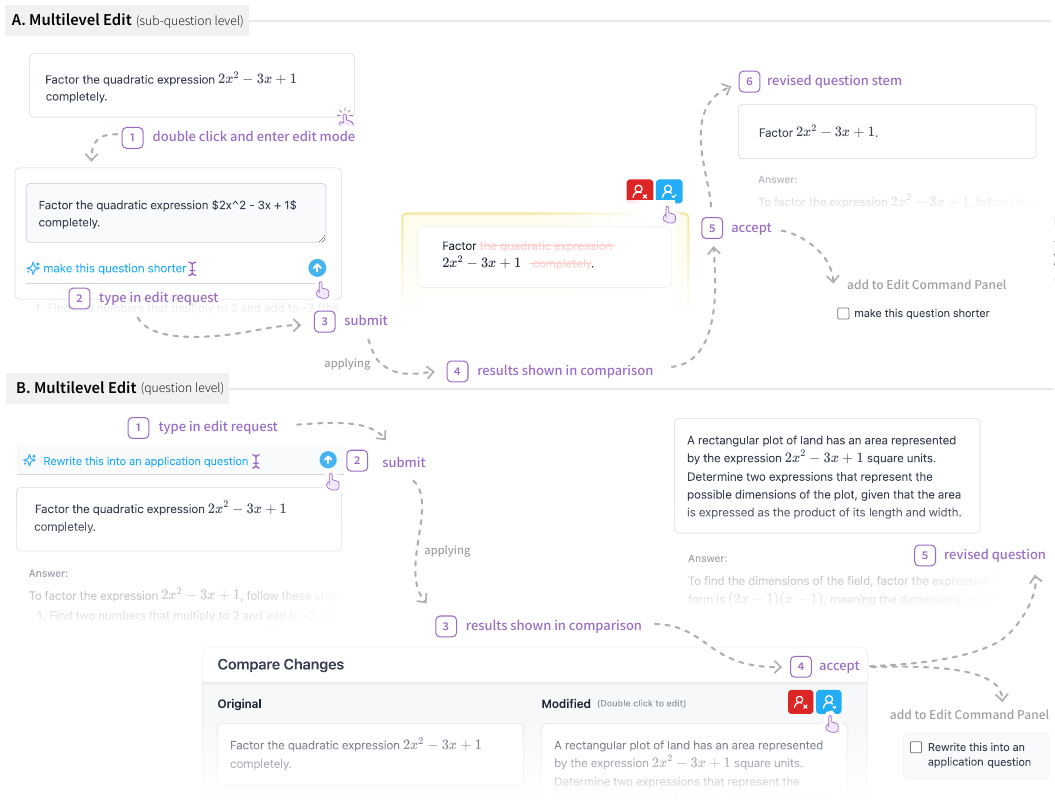}
  \caption{Multilevel edit in \ripplet{}. Teachers can revise either a specific part (top) or the entire question (bottom) with \ai. Differences are shown inline or side-by-side for quick review.}
\label{fig:multi-gran-edit}
\Description{Figure 5: “This figure illustrates multilevel editing in Ripplet, where teachers can revise either a specific part of a question (sub-question level, top) or the entire question (question level, bottom). In Panel A (sub-question level), the teacher begins by double-clicking the question stem ‘Factor the quadratic expression 2x^2 – 3x + 1 completely’ to enter edit mode (Step 1). They type in an edit request such as ‘make this question shorter’ (Step 2) and submit it (Step 3). The system applies the edit and shows results in comparison (Step 4). The teacher can then accept the change (Step 5), which produces a revised question stem ‘Factor 2x^2 – 3x + 1’ (Step 6), and optionally add the edit to their AI Edit Panel for reuse. In Panel B (question level), the teacher types an edit request such as ‘Rewrite this into an application question’ (Step 1) and submits it (Step 2). The system again shows results in comparison (Step 3), here in a side-by-side ‘Compare Changes’ panel. The original question (‘Factor the quadratic expression 2x^2 – 3x + 1 completely’) is displayed next to the modified version (‘A rectangular plot of land has an area represented by the expression 2x^2 – 3x + 1 square units. Determine two expressions that represent the possible dimensions of the plot, given that the area is expressed as the product of its length and width.’). The teacher accepts the change (Step 4), resulting in a revised application-style question (Step 5), which can also be added to the AI Edit Panel. Differences are highlighted inline or side-by-side to support quick review.”}
\vspace{-3ex}
\end{figure*}

\myparagraph{Multilevel Edit.} 
At the sub-question level, double-clicking on a specific part of a question allows teachers to 
prompt an \textsc{llm} to edit just that part (\Cref{fig:multi-gran-edit}A; \DOref{do1}{DO1}). To show the \llm-edited version, \ripplet{} uses an inline difference view that highlights modifications, enabling a focused review of change and quality (\DOref{do3}{DO3}). Teachers can then accept or reject the change. This allows teachers to adapt questions without manually editing content. \refinement~This feature was implemented after the first round of design iteration, where we noticed that many partners wanted to edit only specific parts of the question to prevent unwanted changes in other parts.
At the question level, teachers can edit the entire question using the top prompt bar (\Cref{fig:multi-gran-edit}B). Once the \textsc{llm} completes the request, \ripplet{} displays a side-by-side comparison of the original and \llm-edited versions so that teachers can easily review the change and question quality (\DOref{do3}{DO3}) and decide whether to accept or reject the change. All \llm edit requests are added to the edit command panel for later reuse.\looseness=-10

\myparagraph{Inferred Reusable Edit.} 
Whenever a user makes a manual edit, the system sends the previous and updated versions of the question to an \llm and requests it to infer why this change was made and generate a generalized edit command on the command panel that teachers can later apply to other questions. In \Cref{fig:teaser}, for example, a math teacher edits a question stem to change fractions to integers. The system infers the underlying requirement (\textit{change fractions to integers to reduce difficulty}) and adds it to the command panel (\DOref{do4}{DO4}). \looseness=-10

%

\begin{figure*}
\includegraphics[width=\textwidth]{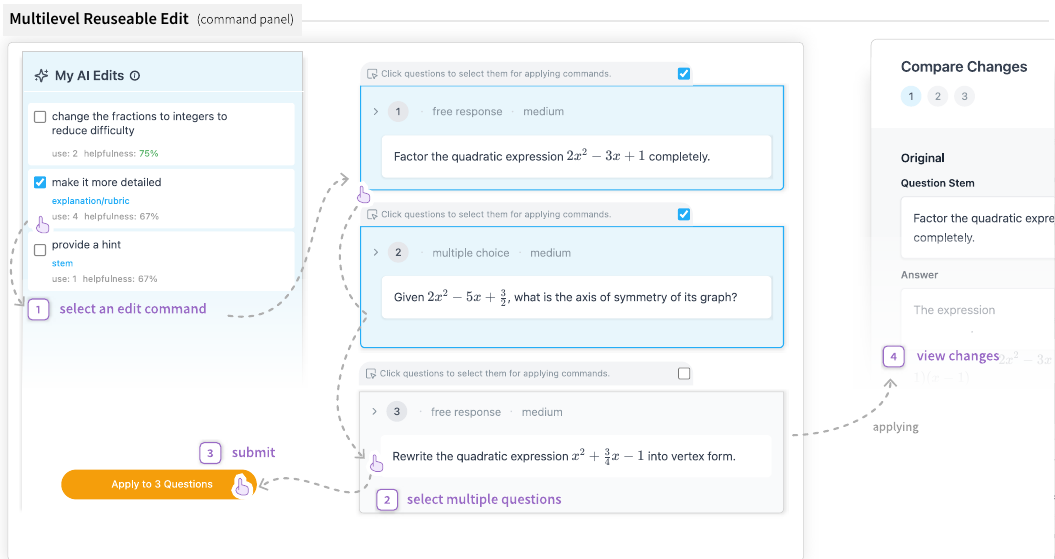}
    \captionof{figure}{Multilevel reusable edits from the command panel. Teachers can select a command and apply it to specific questions. They can review the \ai-generated changes with the original question side by side  and accept or reject each change individually.}
    \label{fig:batch-ai-comparison}
    \Description{Figure 6: “This figure shows multilevel reusable editing from the command panel. On the left, the My AI Edits panel lists reusable edit commands such as ‘change the fractions to integers to reduce difficulty,’ ‘make it more detailed,’ and ‘provide a hint,’ each with helpfulness percentages and usage counts. Step 1 highlights selecting an edit command, here ‘make it more detailed.’ Step 2 shows selecting multiple questions to apply the command. Three questions are displayed: (1) a free-response question ‘Factor the expression 2x^2 – 3x + 1 completely,’ (2) a multiple-choice question ‘Given 2x^2 – 5x + 3/2, what is the axis of symmetry of its graph?,’ and (3) another free-response question ‘Rewrite the quadratic expression x^2 + 3/4x – 1 into vertex form.’ The teacher checks the boxes next to the first two questions to include them. Step 3 shows the submission step with a button labeled ‘Apply to 3 Questions.’ On the right, Step 4 highlights the Compare Changes panel, where AI-generated edits are displayed side by side with the original question for review. Tabs allow switching between multiple questions, and teachers can accept or reject each change individually.”}
\end{figure*}


\myparagraph{Multilevel Reusable Edits.} The edit command panel (\Cref{fig:main_interface}\,\MId) tracks the edits a teacher has made and allows them to apply edits to multiple questions simultaneously, supporting efficient question adaptation (\DOref{do1}{DO1}). This panel automatically stores the multilevel edits a teacher has made and the inferred edit commands from manual edits. Teachers can also create new edit commands from scratch by typing into the text box at the bottom of the panel. Each command has two metrics: \emph{uses} (the number of times the edit has been applied) and \emph{helpfulness} (calculated by updating a beta-distributed prior on helpfulness with the counts of accepted and rejected edits). The panel supports sorting by either metric or keyword search to quickly locate commands.

To apply a command, teachers select it from the panel and choose one or more target questions (\Cref{fig:batch-ai-comparison}). A comparison modal (right side of \Cref{fig:batch-ai-comparison})
allows
the teacher to toggle between questions and accept or reject the \textsc{llm}-generated version individually to ensure question quality (\DOref{do3}{DO3}). 

Initially, we only stored the text of each edit command without storing which part of the question a teacher was editing when that command was created. This meant that the entire question could be changed if a command was reused on another question. For example, if a teacher edited a question stem by asking to ``make this shorter'' then reused the resulting command from the command panel on another question, it would attempt to make every part of that question shorter. During iteration, we observed that teachers wanted to control which parts of a question to edit when applying a command from the panel. \refinement\ We added adjustable tags to commands that specify which
parts of a question it should apply to (\Cref{fig:tags}).



\myparagraph{Reusable Commands As a Way to Enact Requirements.} The reusable commands not only help teachers adapt assessments, but also better \stages{integrate teachers' implicit and explicit requirements} into assessment authoring by tracking and surfacing their edits (\DOref{do4}{DO4}). Initially, we experimented with designs where requirements are reified in the main interface, e.g. as a nested checklist. We imagined the system could help teachers write down 
and mark off items on that list by identifying which question(s) contribute to each requirement. We quickly realized a reified requirement checklist does not fit the implicit and entangled nature of assessment requirements. Some requirements are difficult to articulate up front, such as using only positive numbers in a question about factoring so that students can focus specifically on factoring. Other requirements are entangled: for example, when teachers adjust question difficulty, they might simultaneously change the coverage and progression of topics in the assessment.
It would be difficult to isolate and articulate all such requirements explicitly.
However, we noticed that teachers often discover requirements as they work on specific questions. We designed multilevel reusable edits to capitalize on this behavior by eliciting, tracking, and surfacing a teacher's requirements implicitly from their edits, without imposing the burden of writing down explicit requirements.\looseness=-1

\myparagraph{Managing the Command Panel with Similarity Checker.}
Teachers often make many edits to an assessment. 
During design iteration, this could generate many edit commands and quickly overwhelm the command panel, reducing its usefulness. While teachers could use the search and delete functions to manage the panel, this requires additional effort. 
However, many edits are similar. For example, P3 repeatedly removed extra words to keep question stems concise, and P12 repeatedly used \ai edits like ``add a hint'' or ``make the question easier by giving additional information''. \refinement\ To reduce the length of the command panel, we implemented an \textsc{llm}-based similarity checker. When a new command is generated, 
it is passed to an \textsc{llm} along with the current list of commands on the panel to judge if the new command is sufficiently different from all existing commands. To ensure consistency, we collected common examples of similar commands and used few-shot prompting for the checker's prompt template to help define command similarity (e.g., ``make it shorter'' and ``make more concise'' are similar). The similarity checker significantly reduced the number of repeated commands on the panel and made the panel more usable for teachers.

\subsection{Cross-Assessment Operations and Supports}\label{subsec:crossassess}
\ripplet{} also provides tools that operate across assessments to help teachers reuse and adapt materials (\DOref{do1}{DO1}).

\myparagraph{Sharing Edit Commands across Assessments.}
Initially, the edit commands were only preserved within an assessment. 
Multiple partners, including P1 and P12 who used \ripplet{} extensively, indicated that they needed to apply similar edit commands across 
assessments 
and even across courses. Therefore, \refinement\
we made each teacher's
edit commands available across their assessments 
to help them adapt other assessments without re-specifying commands (\DOref{do1}{DO1}).
\myparagraph{Importing Questions from Other Assessments.} To support reusing high-quality content (\DOref{do1}{DO1}), \ripplet{} allows teachers to add questions from previous assessments in the system via the search and import bar (\Cref{fig:main_interface}\,\MIb). Teachers can 
\stages{search and filter past} \stages{assessment questions} by a variety of criteria, such as class, topic, format, and difficulty, then view full question details and directly insert questions into the current assessment.

\myparagraph{Equation Authoring and Syntax Highlighting.}
In our formative interviews, mathematics and science teachers described significant challenges in creating, editing, and formatting expressions and equations using existing tools like Microsoft Word. For example, P12 struggled to create a cube-root symbol, so they used the square-root symbol available in their software and hand-wrote the ``3'' on every copy of the test. To support a wide range of symbols and display them properly, \ripplet{} enables users to render and author expressions and equations using \LaTeX{} notation. Recognizing that teachers typically lack familiarity with \LaTeX{},Ripplet allows teachers to generate and edit \LaTeX{} symbols with an \textsc{llm} and automatically displays a rendered preview whenever mathematical expressions or markdown blocks are detected. 




To support computer science teachers, during the second round of design iteration, we implemented syntax highlighting for all major programming languages using the same tab interface for displaying rendered results (\raisebox{-6pt}{\includegraphics[height=17pt]{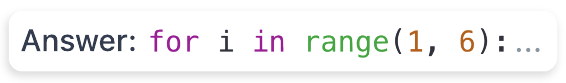}}). 

\begin{figure*}
    \vspace*{-10pt}
    \includegraphics[width=\textwidth]{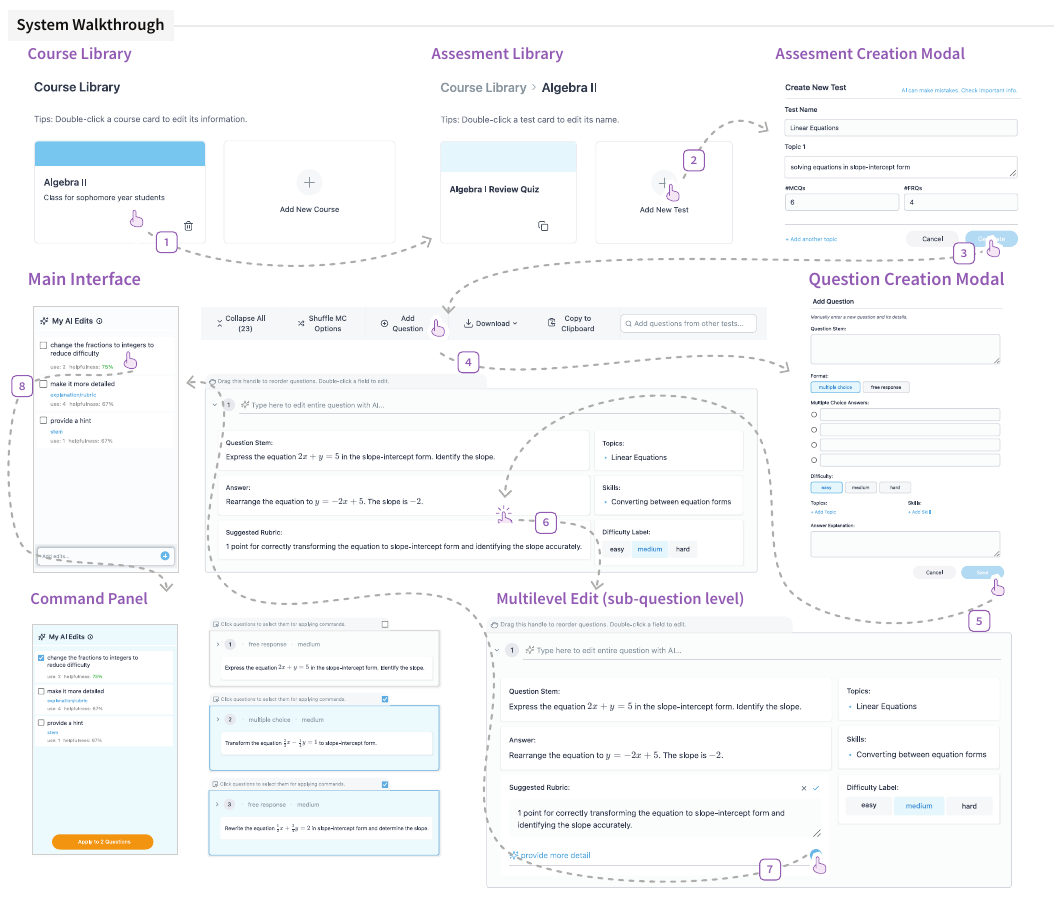}
    \vspace*{-20pt}
    \caption{\textbf{System walkthrough of Ripplet.} Ms. Ripley starts assessment creation by generating from a topic. She then manually writes a question and uses \ai to edit it. Finally, she notices that some questions are too hard, so she applies an AI edit command to edit multiple questions at once. She reviews the revisions and accepts the suitable changes to finalize the test.}
    \vspace*{-10pt}
    \label{fig:walkthrough}
    \Description{Figure 8: “This figure is a system walkthrough of Ripplet showing how a teacher creates and edits an assessment. The flow is illustrated with numbered steps connected by arrows across several interface screenshots. Step 1: In the Course Library, the user sees course cards such as “Algebra II” and can add a new course. Step 2: In the Assessment Library inside Algebra II, the user can view existing assessments like “Algebra II Review Quiz” and click “Add New Test.” Step 3: In the Assessment Creation Modal, the user enters a test name such as “Linear Equations,” specifies the topic (e.g., “Writing an equation in slope-intercept form”), and can add multiple topics. Step 4: The Main Interface displays the test in detail. It includes a question stem such as “Express the equation 2y – 5x = 5 in the slope-intercept form. Identify the slope,” answer fields (showing a rearranged equation with slope = 2), suggested rubric text, topics, skills, and a difficulty label (easy, medium, hard). Step 5: The Question Creation Modal is used to add new questions, with fields for question stem, answers, explanations, and metadata like topics and skills. Step 6: In the Main Interface, a multilevel edit view shows the sub-question level where users can refine the stem, answers, and rubric. Step 7: The user can also trigger multilevel edits with AI, shown with highlighted fields for editing the stem or generating alternatives. Step 8: On the left-hand side, the Command Panel called “My AI Edits” lists edit commands such as “change the fractions to integers” or “provide a hint,” each tagged with scope (e.g., stem, explanation, rubric) and a helpfulness percentage. The caption explains: Ms. Ripley begins in the course library, navigates into the assessment library, and creates a new assessment through the test creation modal. She edits it in the main interface, reviews and edits questions manually or with AI, and finally exports the test as a PDF.”}
\end{figure*}

\subsection{Ripplet Walk-through}\label{subsec:walkthrough}
To demonstrate how Ripplet supports
assessment authoring and realize the conceptual model, we walk through a fictional scenario where a teacher, Ms.~Ripley, creates an assessment (\Cref{fig:walkthrough}). We highlight the connections to the conceptual model by marking the \inputs{inputs} and \stages{stages}.

Ms.~Ripley is teaching an Algebra~II course that has a test coming up on linear equations.
She logs in to \ripplet{}, enters her course library and clicks on her Algebra II course. She enters the assessment library and begins creating her test. Ms.~Ripley sees several ways to do this and chooses  ``Generate from Topic'' because she 
has specific topics in mind:
\inputs{solving equations in slope-intercept form} and \inputs{converting them to standard form}. She also has a rough idea of how many multiple-choice and free-response questions a balanced test should have to take up the full class period.

After she clicks "Generate", Ms. Ripley enters the main interface with the assessment. After quickly \stages{reviewing} the questions, she decides to \stages{add additional questions} to \stages{cover a missing topic}. This opens a modal with the same question creation methods she saw before. This time she \stages{writes a new question} on linear equations from scratch. Now that her test satisfies her length and topic coverage requirements, Ms. Ripley carefully \stages{reviews} and \stages{revises each question}. She notices that the suggested rubric for a free-response question is not detailed enough, so she double clicks on it and asks \ai to provide more detail. She \stages{checks the \ai's output} and notices that the new rubric is too detailed, so she rejects the change and edits it manually. She then notices that some questions may be \inputs{too hard for her students} because the solutions involve fractions, which her students have been struggling with. For this test, she only wants to assess students' ability to work with equations of a line, not their understanding of fractions, so she reuses the command ``change the fractions to integers to reduce difficulty'' in the command panel and applies it to all corresponding questions at once. She reviews the changes and accepts the suitable ones. Once satisfied, she downloads it as a PDF to hand out in class.

\subsection{System Implementation}
We use OpenAI's GPT-4o for Ripplet's \textsc{llm}-related features and specifically GPT-4o with vision for generating questions from curriculum guides and importing questions, as it can extract special symbols and expressions from images of documents. Some prompt templates benefited from collecting examples from codesign partners. For example, we used common examples of similar commands to revise the prompt template of the similarity checker. \looseness=-1

In the database \footnote{We provide the database design documentation in supplemental materials.}, we store the full edit history
of questions for version control. We also store the full history of edit command usage to calculate the helpfulness and use metrics. In the interface, we render mathematical equations for in-browser display and copy-to-clipboard using \verb|remark-math| \cite{remark-math} and \verb|rehype-katex| \cite{rehype-katex}, while for PDF rendering we use \verb|KaTeX| \cite{katex}. Similarly, for syntax highlighting, we use \verb|react-syntax-highlighter| \cite{react-syntax-highlighter} for in-browser display and copy, while \verb|prismjs| \cite{prismjs} is used for rendering PDFs. 
\charles{include database design in supp to please R3?}

\section{Codesign Phase III: Independent Use and Longitudinal Observations}\label{sec:phase3} 
Our codesign process provided a unique opportunity to track \ripplet{}'s use across two complementary dimensions: independent adoption and longitudinal engagement. Seven teachers used \ripplet{} independently to create and administer assessments in their classrooms, giving us insight into how well Ripplet functioned in organic settings and captured the the workflows in our conceptual model.
In addition, because many of the same teachers participated in multiple rounds of codesign sessions and continued corresponding between sessions, we were able to observe how their practices and preferences evolved over several months.

\subsection{Independent Classroom Adoption}\label{sec:independent_classroom}
Following the second round of design iteration, seven teachers used \ripplet{} to create assessments and administered them to their students. They created a variety of assessments—from official final exams for state-required classes to practice tests for dual-credit community college courses.

\myparagraph{\ripplet{} Integrated into Teachers' Existing Workflows.} Teachers' independent use revealed that \ripplet{} fits naturally into a variety of existing workflows by supporting different \inputs{inputs} and \stages{stages} in the conceptual model. For example, P9 \stages{found high-quality \textsc{ap}} \stages{questions}, imported them, and used the ``Generate Similar'' and \ai editing features to create a review packet, while P8 uploaded \inputs{curriculum standards} to generate a mock exam for students in their health careers course and used a mix of manual editing and the command panel to \stages{reduce the reading level} of some questions. P1 started with topic-based generation to target specific areas; some questions \stages{inspired} \stages{and reminded them to cover a different topic} and blend multiple topics together into one question, so they created new ones both manually and with \ai. Finally, they \stages{reordered the questions so} \stages{that they progress in difficulty}. By supporting diverse question generation, adapting questions at multiple levels, and restructuring assessments, \ripplet{} allowed teachers to build on their materials and navigate the iterative dual authoring process in the conceptual model without abandoning their established ways of working. 

\myparagraph{Time Savings Enabled Teachers to Create Materials They Could Not Before.} Teachers expressed that \ripplet{} enabled them to create assessments they previously lacked the time or resources to produce. P8, for instance, created a 28-question practice final exam in just 30 minutes by using curriculum standards and applying reusable edit commands to \stages{scaffold questions and adjust language} \stages{level}—a task that would normally have taken them over five hours. They emphasized that the efficiency made the creation of more formative assessments possible: ``\inlinequote{I just wouldn’t have made it otherwise.}'' Similarly, P9 explained that without \ripplet{}, they would have had to purchase a commercial review packet, but instead they were able to build one themselves quickly and give it to their students. Teachers noted that these new assessments had clear downstream benefits for students: P8’s class treated the mock exam as seriously as their actual final and learned a lot by working through the mock exam, and P9’s students reported that the review packet was helpful in preparing for the final exam. These accounts underscore \ripplet{}'s educational impact of not only saving teachers’ time but also expanding the opportunities they could offer their students.

\myparagraph{Multilevel Organization Eased Assessment Refinement.} Teachers also praised how \ripplet{}’s structured interface supported the stages of assessment authoring, particularly restructuring assessments and revising questions. The system’s card-based question display and menu bar helped them \stages{reorder questions} and \stages{shuffle} \stages{multiple-choice options}—tasks that were cumbersome in tools like Google Docs, which required manually cutting and pasting text. P12 described the organizational structure as ``\inlinequote{the huge benefit with \ripplet{} ... each question is its own element, and it’s easy to make localized changes.}'' Being able to edit specific parts of questions with \ai further supported this process by allowing teachers to \stages{refine multiple-choice distractors} or \stages{rephrase question stems} without disturbing the other parts of questions <P3>. The multilevel organization gives teachers more control over how their assessments were assembled, supporting evaluation and restructuring while reducing the tedium of manual formatting.

\subsection{Behavioral Changes in Assessment Authoring}\label{subsec:behavioral}
\myparagraph{Evolving Ownership: from Generation to Curation.} As teachers engaged with \ripplet{} over time, their sense of ownership and role in assessment design shifted. Early in the codesign process, our partners spent more time generating questions with \textsc{llm}-based methods. While they emphasized the need to reuse and adapt prior materials, they did not utilize many features designed for that purpose (\eg importing assessments, command panel). As a result, many perceived \ai as the ``real'' author and themselves as proofreaders, often questioning whether they could truly claim ownership of the assessments. With repeated use, however, their behavior and perception evolved. Teachers routinely imported questions from their own past tests, modified them for new contexts, and wove them into new assessments. P12 highlighted the efficiency of being able to ``\inlinequote{add questions from other tests and filter them}.'' Similarly, the command panel for reusing edit commands—initially underexplored—grew into a valued tool. In the first round, some overlooked the command panel entirely; by the second, they praised it as a timesaver that improved productivity <P4, 8>. Teachers also grew comfortable deleting low-quality questions, rejecting weak \ai suggestions, and refining assessments manually. Teachers increasingly recognized themselves as the primary creators. P9 explained that at first they felt ``\inlinequote{the \ai did most of the work,}'' but later came to see that they were ``\inlinequote{the one deciding what to keep, what to change, and what made sense for my students.}'' In addition to teachers' growing familiarity with \ai and their development of more effective prompting strategies, several features of the system likely contributed to this transition, such as the side-by-side comparison modal and inline difference view of \ai editing, which allows teachers to control the quality (\DOref{do3}{DO3}) of assessments by reviewing, accepting, or rejecting \ai edits easily.

\myparagraph{Becoming More Reflective by Writing and Seeing Commands.} Teachers also reported that \ripplet{} made them think more deliberately about assessment design and question quality. P1 explained that using \ai to adapt questions forces them to think more critically about what it is that makes a question difficult. Likewise, P3 reflected that their usage of \ripplet{} encouraged them to slow down and become ``\inlinequote{more deliberate in what I do, instead of just going through the motions.}'' Similarly, P12 expressed that seeing the commands showing up on the command panel helped them \stages{understand their own requirements} and reflect on what needed to be done to design a high-quality assessment (\DOref{do4}{DO4}). By inferring and surfacing teachers’ requirements as reusable edit commands, the system helped teachers approach assessment authoring in a more systematic and reflective way.

\myparagraph{Extending \ripplet{}’s Applications in Teaching.} Extended engagement also reshaped how teachers envisioned their professional practice. P8 described wanting to use \ripplet{} on demand during office hours to generate practice questions, while others imagined using it for multilingual or remedial contexts (\ripplet{} supports the display of text in multiple languages). Teachers also highlighted time saved outside of assessment authoring: P3 and others noted that \ripplet{}’s answer keys and explanations allowed them to hand these directly to students, reducing the need for lengthy office hours. P1 explained that students can self-diagnose first instead of them hosting a multi-hour session to explain the answers. These reflections show that teachers came to see \ripplet{} not only as an assessment authoring tool but also as part of a broader ecosystem of teaching, learning, and collaboration.
\section{Controlled User Study}\label{sec:user_study}
While our codesign process showed how Ripplet supported diverse workflows in the conceptual model and how teachers engaged with it over time,
it involved codesign partners who were deeply familiar with the system. To evaluate \ripplet{} on a sample of teachers less familiar with the system or invested in its design, we conducted a controlled, within-subjects study with 15 other teachers. Each teacher created assessments in two conditions: once with their current practice (control) and once with \ripplet{}. This design allows us to make a direct comparison between \ripplet{} and their current practices. It can also provide evidence on whether Ripplet's benefits and integration to teachers' workflows generalize beyond our codesign partners.



\subsection{Participants}

\aboverulesep=0ex
\belowrulesep=0ex
\setlength\arrayrulewidth{.2pt}
\bgroup
\begin{table}[!t]
\centering
\caption[]{Information about teachers in controlled user study.}
\small
\vspace*{-8pt}
\def\arraystretch{1}%
\label{tab:user-study}
\begin{tabular}{>{\arraybackslash}m{0.02\textwidth}
                m{0.03\textwidth}
                m{0.03\textwidth}
                m{0.03\textwidth}
                m{0.06\textwidth}
                m{0.16\textwidth}}
\midrule\\[-6pt]
\textbf{Id}   & 
\textbf{Sex} & 
\textbf{Years} & 
\textbf{State}  & 
\textbf{School}  & 
\textbf{Subjects} \\ [2pt]
U1  & F & 19 & OH & Private  & Geometry, Pre-Algebra   \\
U2  & M & 8  & OH & Public  & Chemistry    \\
U3  & F & 18 & VA & Private & Economics, Human Geography, Business          \\
U4  & F & 30 & OH & Public  & Biology, Anatomy          \\
U5  & F & 1  & IL & Public  & Biology, Physiology          \\
U6  & F & 8  & TX & Public  & Biology          \\
U7  & M & 31 & OH & Public  & Middle School History          \\
U8  & M & 13 & TX & Public  & Economics, Financial Literacy  \\
U9  & F & 27 & OH & Private & History, Psychology   \\
U10 & F & 7  & OH & Public  & Pre-Algebra, Algebra I, Geometry        \\
U11 & F & 5  & OH & Public  & Algebra I    \\
U12 & F & 6  & OH & Public  & Algebra II, Statistics \\
U13 & F & 20 & OH & Private & Middle School Math    \\
U14 & F & 29 & OH & Public  & Financial Literacy \\
U15 & M & 15 & IL & Public  & Computer Science \\
\midrule
\end{tabular}
\vspace*{-10pt}
\end{table}
\egroup

\aboverulesep=0ex
\belowrulesep=0ex
\setlength\arrayrulewidth{.2pt}
\bgroup
\begin{table*}[h]
\centering
\caption[]{Survey items grouped by five areas. Participants were asked to rate each item on a scale from 1 to 10.\footnotemark}
\small
\vspace*{-8pt}
\def\arraystretch{1}%
\label{tab:survey-questions}
\begin{tabular}{>{\raggedright\arraybackslash}p{0.18\textwidth}
                >{\raggedright\arraybackslash}p{0.72\textwidth}}
\midrule\\[-6pt]
\textbf{Area} & \textbf{Survey Item}: (highly disagree [1] - highly agree [10])\\ [2pt]

Enjoyment & I would be happy to use this way of creating assessments on a regular basis. \\
          & I enjoyed using this way of creating assessments. \\[4pt]

Exploration & It was easy for me to explore many different ideas, options, designs, or outcomes, using this way of creating assessments. \\
            & This way of creating assessments was helpful in allowing me to track different ideas, outcomes, or possibilities. \\[4pt]

Results Worth Effort & I was satisfied with what I got out of this way of creating assessments. \\
                     & What I was able to produce was worth the effort I had to exert to produce it. \\[4pt]

Perceived Control & I felt I had a say in the assessment creation process. \\
                  & I was able to influence the assessment creation process. \\[4pt]

Assessment Quality & The assessment I created can measure my students’ skills and knowledge or help them improve. \\
                   & The assessment questions are worded clearly. \\
                   & The assessment is adequately difficult for my students. \\

\midrule
\end{tabular}
\vspace*{-10pt}
\end{table*}
\egroup
\footnotetext{Although the 1–10 scale does not have a neutral midpoint, its granularity allows sufficient gradation in responses, and because our analysis focuses on differences across conditions, the absence of an explicit neutral point is unlikely to affect the results.}

We asked our codesign partners to reach out to their teacher networks through mailing lists and Facebook groups. \Cref{tab:user-study} shows the 15 teachers (11F, 4M) we recruited from 13 schools in 4 states, using the same criteria as codesign (see \Cref{sec:codesign_participants}). The group included 13 high school and 2 middle school teachers across different subject areas: mathematics and computer science (6), natural sciences (4), and social studies (5). Their teaching experience ranged from 1 to 31 years. All participants completed a pre-survey that included questions about their experience and comfort with using \ai tools. Some had never used \ai\ at work and self-identified as having ``no ability'' in using \ai <U14>, while others identified as ``advanced'' users and had used tools such as ChatGPT to create and adapt assessments frequently <U2>. The pre-survey also asks each teacher to list the courses they will teach next semester and two chapters or units in each course that are similar in length and difficulty. None of the participants had previously used Ripplet or seen its interface; their only prior exposure was the recruitment message indicating that Ripplet is an \textsc{ai}-powered tool for authoring assessments. This study was \textsc{irb}-approved, and each participant received a total compensation of 195 \textsc{usd}. 

\subsection{Procedure}
The study involved two Zoom meetings, each followed by an asynchronous task that teachers conducted and recorded themselves. To counterbalance potential order effects, we randomly assigned participants to either a control-first or \ripplet{}-first group, then randomly picked a course they listed in the pre-survey and randomly assigned two units from that course to the two conditions.

In the control-first group, participants began with a 15-minute Zoom session where we introduced the study and outlined the first asynchronous task. In this task, we asked teachers to create an assessment for a unit of a course that they would teach the next semester, using their current method of authoring assessments. While working on the task, they recorded their screens and narrated their process by thinking aloud. Afterward, they completed a survey (\Cref{subsec:survey_design}) reflecting on their experience. The second Zoom meeting (60 minutes) began with a 10-minute tutorial video introducing the \ripplet{} tool followed by 40 minutes of hands-on exploration. During this time, participants experimented with \ripplet{} and received support from the research team as needed. In the last 10 minutes, we introduced the second asynchronous task: creating an assessment using \ripplet{} for the unit assigned to the \ripplet{} condition. As with the first task, the participants recorded their screens, narrated their process, and completed a post-task survey. In the \ripplet{}-first group, participants followed the same set of activities and tasks, but in reverse order. 

\subsection{Survey Design}\label{subsec:survey_design}
We asked participants to evaluate \ripplet{} and their current practices in five areas through an 11-item survey (see \Cref{tab:survey-questions}). 
As assessment authoring is a open-ended creative task, we first adapted the Creativity Support Index (\textsc{csi}), which offers a meaningful and validated way to assess how well a tool supports creative work \cite{csi}. From \textsc{csi}, we selected the areas applicable to evaluating our system: ``Enjoyment'', ``Exploration'', and ``Results Worth Effort''. Because we need to administer the survey in both the control and \ripplet{} conditions, we rephrased the \textsc{csi} statements that read ``the/this system or tool'' to ``this way of creating assessments''. To understand users' sense of agency and control in a system involving \llm{}s, we measured perceived control with two questions from Lee et al.~\cite{perceived-control} and adapted them by changing ``content editing'' to ``assessment creation'' to suit our setting. Finally, to understand the educational impact of \ripplet{}, we added three questions regarding the quality of assessments. We adapted two questions from Cui \etal~\cite{cui2025vila} for measuring the relevance and clarity of assessments, and we added another question on the difficulty of the assessments. 
We asked teachers to self-evaluate assessment quality rather than rely on external raters, because what constitutes a ``high-quality'' assessment differs across school settings and student cohorts even within the same course, making it difficult to define a universal, objective standard \cite{mcmillan1999establishing,swaffield2008unlocking}. \remove{While external ratings are possible, doing so would require recruiting subject-specific experts across the wide range of disciplines our users teach and providing them with sufficient contextual information about each class to support fair judgments. Coordinating such a large group of raters and facilitating communication between raters and teachers would be very resource-intensive. After weighing these considerations, we determined that teacher self-evaluation offers a reasonable proxy for assessment quality at this stage of the design process.}
We also invited five codesign partners (P1, 3, 7, 12, 13) for feedback on the survey. P7 expressed that only those who know the students’ abilities and struggles can fairly judge assessment quality, supporting our decision to use self-evaluation. We finalized the survey after codesign partners confirmed that the questions capture the most important aspects of assessment quality. 

\subsection{Quantitative Comparison}
To compare participants' authoring experiences and the quality of assessments from using \ripplet{} vs. their current practices (control), we compared their ratings in each of the five areas on the survey.
\myparagraph{Analysis.}
First, based on the guidance for analyzing Likert items, we averaged ratings for items within each area for each participant and condition \cite{50years,likert-scales}. For each area, we then compared \ripplet{} vs. control using paired-samples $t$-tests on the within-participant mean differences (\ripplet{} $-$ control). To account for the five parallel tests, we controlled the false discovery rate using the Benjamini–Hochberg (\textsc{bh}) procedure ($\alpha=0.05$), and we report the \textsc{bh}-adjusted $p$ value, mean difference, and its 95\% CI for each of the five areas. We chose the paired $t$-test because it is high-power and sufficiently robust to modest assumption violations \cite{likert-scales,Stonehouse01021998,ban_of_p}.\footnote{$t$-tests can be performed on low-number-of-item Likert scales because parametric statistics are robust with respect to Likert being ordinal \cite{likert-scales}.} This analysis plan was pre-registered on \textsc{osf}.\footnote{\textsc{osf} link: \url{https://osf.io/j8zey}
}

\begin{figure*}
    \includegraphics[width=\textwidth]{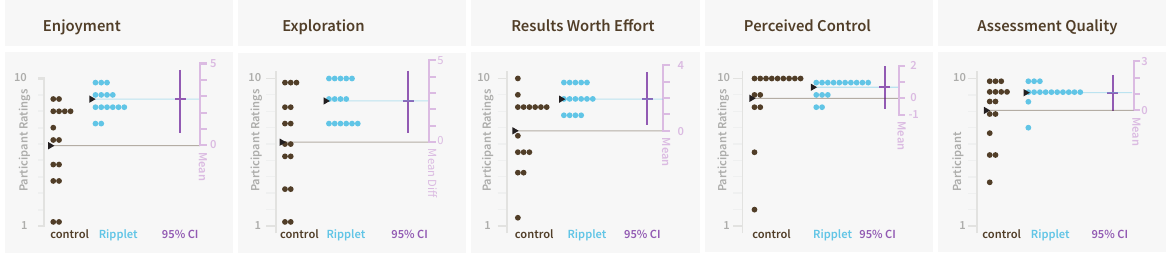}
    \caption{Distribution of teachers' ratings for each of the five areas in both conditions, with mean differences (\ripplet{} $-$ control) and their 95\% CIs from the paired $t$-tests.}
    \label{fig:quant-results}
    \Description{Figure 9: “This figure shows the distribution of teachers’ ratings for five areas (Enjoyment, Exploration, Results Worth Effort, Perceived Control, and Assessment Quality), comparing control and Ripplet conditions. Each panel is a dot plot of individual participant ratings from 1 to 10 (y-axis) with control ratings shown as grey dots and Ripplet ratings as blue dots. To the right of each panel is a purple mean difference plot showing the average difference (Ripplet minus control) with 95\% confidence intervals from paired t-tests. Overall patterns: - In the Enjoyment panel, Ripplet ratings cluster higher than control ratings, and the mean difference is positive with confidence intervals above zero. - In the Exploration panel, Ripplet ratings cluster higher than control ratings, and the mean difference is positive with confidence intervals above zero. - In Results Worth Effort, Ripplet ratings cluster higher than control ratings, and the mean difference is positive with confidence intervals above zero. - In Perceived Control, the two conditions are similar, with little mean difference. - In Assessment Quality, Ripplet ratings cluster higher than control ratings, and the mean difference is positive with confidence intervals above zero. The figure summarizes both raw rating distributions and mean paired differences with their confidence intervals.”}
\end{figure*}

\myparagraph{Results.} \Cref{fig:quant-results} shows the distributions of participants' ratings for the five areas in both conditions, as well as the mean and the confidence interval of the difference in ratings (\ripplet{} $-$ control). \textbf{\ripplet{} shows significant improvement over control in four areas}: enjoyment ($\mu=+2.70$ with 95\% CI of $[0.86,4.54]$, $p=0.003$), exploration ($\mu=+2.43$ $[0.56,4.31]$, $p=0.012$), results worth effort ($\mu=+1.93$ $[0.33,3.53]$, $p=0.012$), and assessment quality ($\mu=+1.11$ $[0.10,2.11]$, $p=0.032$). Ratings for perceived control are similarly high across conditions, possibly because teachers in both settings could manually create and edit questions. 

\subsection{Qualitative Findings}\label{sec:qual-findings}

In addition to rating both conditions, participants answered two open-ended questions: (1) which features or functionalities they found most helpful, and (2) the difficulties they encountered. We also coded\footnote{We coded the videos by indicating the specific features users used in Ripplet. The codes are provided in supplemental materials.} participants’ self-recorded usage of Ripplet to understand their workflows and features used. 
Their responses and the patterns we observed in their self-recordings corroborate our observations from the codesign process (\Cref{sec:phase3}). Many users began by importing \inputs{curriculum standards} or \inputs{past assessments} then used a mix of manual edits and \ai edits to \stages{adapt questions} and \stages{reorganize} \stages{assessments}. Participants highlighted that multilevel reusable edits were efficient for revising assessments <U3, 5, 8, 10, 12>. For example, U8 applied the command ``reword this question so that a non-native speaker could understand it'' across several questions to accommodate students who needed language support. Teachers also valued features such as generating similar questions, which allowed them to reuse high-quality materials <U2, 13>, and shuffling MC options <U9> to \stages{restructure assessments}. U8 noted that \ripplet{} would enable them to create more formative assessments for the upcoming school year—assessments they otherwise would not have had time to prepare.

At the same time, several math teachers mentioned their unfamiliarity with \LaTeX\ and often had to ask \ai to correct related issues <U11-13>. U13 expressed confusion that similar prompts sometimes produced different results, suggesting that understanding the uncertain nature of \llm{}s and adjusting expectations could take time. Teachers also expressed a desire for support in generating questions with visual representations <U5, 8>, suggesting an important area for future work on assessment authoring systems.



\section{Discussion, Limitations, and Future Directions}

\edit{We discuss limitations of our work and directions for future research. We reflect on how characteristics of our teacher cohort shaped the conceptual model and system use; we discuss how uneven risks of AI reliance affect different teachers, motivating the need for safeguards and supports tailored to these differences. Finally, we propose deeper evaluations for multilevel reusable interactions and reflect on the lessons learned from sustaining a long-term codesign process with teachers.}

\subsection{Variations in Teacher Backgrounds Affect Ripplet Usage and Generalizability of Conceptual Model}

Because our codesign partners were largely \textsc{stem} teachers and experienced teachers \edit{(\Cref{sec:codesign_participants})}, many of the behaviors, challenges, and workflows that informed our conceptual model and shaped our system design could reflect practices more common in math and science assessment authoring as well as the habits of veteran teachers with well-developed materials. \edit{Prior work shows that \textsc{stem} and humanities teachers often have different instructional priorities and needs \cite{Riahi_2025}, a distinction that also surfaced in our sample: }\remove{For example,} \textsc{stem} teachers often invested much effort in verifying the accuracy of the answer keys and tweaking numerical details, making \stages{reviewing assessments} and \stages{adapting questions} critical stages in the conceptual model. On the other hand, the two social studies teachers (P2, 13) created more free-response questions and did not spend significant time scrutinizing the answers, shifting the burden of verifying answers to grading rather than authoring. 
New teachers in their first or second year and those from humanities and social studies were underrepresented, limiting our ability to observe how their distinct disciplinary practices and early-career needs might surface different inputs, stages, or transitions within the conceptual model. As a result, the generalizability of our conceptual model may be constrained by the makeup of our codesign sample. Future work could expand the model by recruiting a more balanced set of teachers across subject areas and years of experience to support the full diversity of K–12 teaching contexts.


\subsection{\edit{Mitigating Uneven Risks of AI Reliance Needs Carefully Designed Safeguards}}
\edit{
\edit{Given teachers’ concerns around fairness and safety in \ai-generated content \cite{yin2025}, it is important to consider how variations in teacher backgrounds may produce varied risks.} Teachers who are already stretched for time and materials may be particularly vulnerable to over relying on \ai outputs without the capacity to carefully review them. In contrast, teachers with more experience, training, or institutional support may be better positioned to reuse high-quality content, recognize hallucinations, and adjust their use accordingly. Safeguards and training should be implemented to mitigate these risks.
 Future systems could embed proactive and explainable support mechanisms that help teachers verify, interpret, and improve \ai-generated content as part of their natural workflow. For instance, rather than merely flagging the uncertainty of \llm-generated content, systems could use non-\llm-based verification pipelines to retrieve authoritative references or validated assessment banks to cross-check for correctness \edit{\cite{Maheshwari2025}.}

Another possibility is to make \ai reasoning more transparent and inspectable. Instead of treating the model's internal process as a black box, tools could display their reasoning process or confidence explanations that allow teachers to interrogate the rationale behind generated content \edit{\cite{Lim2009, Kaur2020}.} For instance, systems could show the inferred learning objective, reasoning steps, or aligned curriculum tags that led to its output to give teachers concrete points for critique or correction. Such features could turn validation from a cognitive burden into an interactive process of sense making. For teachers who coauthor assessments with colleagues, systems could also explore collaborative verification workflows. Rather than asking each teacher to independently judge \ai outputs, they could allow peer co-review or lightweight consensus-building interfaces that pool teacher feedback on the generated assessments. Such distributed review processes could surface common errors and strengthen the collective reliability of \ai-generated materials across teachers and schools. Designing for different risks means acknowledging that while \ai can be useful for some teachers, it also carries uneven vulnerabilities. Building effective assessment authoring tools therefore requires equipping users—especially those under the most constraints—with support that match their needs.
}

\subsection{Multilevel Reusable Interactions Demonstrate Promise but Require Deeper Evaluation}

Evidence from codesign and the user study suggests that Ripplet’s multilevel reusable interactions support the iterative dual process in our conceptual model \edit{(\Cref{subsec:conceptual_model} and \Cref{subsec:multi-reuse-edits})}. Teachers used reusable commands to revise assessments, and several teachers (e.g., P8, U8) noted that this feature saved substantial time during assessment authoring \edit{(\Cref{sec:qual-findings} and \Cref{subsec:behavioral})}.
However, we did not obtain objective measures of the effectiveness of multilevel reusable interactions \edit{(\Cref{subsec:survey_design})}. Future work could include controlled A/B testing studies comparing authoring workflows with and without the command panel. Such experiments could quantify changes in authoring time, assessment quality, and teachers’ experiences. Longitudinal studies could also potentially reveal whether reusable edits become more valuable as teachers accumulate a personalized library of commands over time. 
Finally, future work could investigate what types of edits multilevel reusable interactions are particularly well-suited for and where they fall short. Classifying the strengths, limitations, and failure modes of reusable edits
could inform new forms of interaction designs that help teachers understand, trust, and more effectively reuse these commands. Such insights would also guide improvements to the underlying interaction paradigm, ensuring that multilevel reusable edits genuinely support the breadth of assessment authoring tasks teachers perform.\looseness=-10

\subsection{Effective Codesign Requires Aligning with Teachers' Schedule and Supporting their Growth}
Engaging teachers in a seven-month codesign process across two semesters proved invaluable for developing \ripplet{}, and it also surfaced lessons about both the opportunities and challenges of sustained collaboration. Many teachers began with limited prior exposure to \ai tools. Through the codesign process, they developed a better understanding of the probabilistic nature of these systems and learned strategies to frame prompts and manage expectations. In this way, codesign was not just about providing feedback on \ripplet{}, but also an opportunity for teachers themselves to build new knowledge and skills around working with \ai.

At the same time, sustaining participation was demanding. Longitudinal codesign required repeated follow-ups and building trust and relationships. Teachers’ workloads left little time to devote to research activities, and one partner remarked that they wished they had ``\inlinequote{more time to play around with the tool outside \emph{[}of the scheduled Zoom sessions\emph{]}.}” In addition, we timed Phase II to span a semester and Phase III to be when teachers are writing final exams so that teachers could meaningfully test \ripplet{} during periods when they were actively teaching and developing assessments \edit{(\Cref{fig:codesign-overview})}. It is important to not only create intentional opportunities for open-ended exploration outside structured sessions, but also carefully align system design and evaluation periods with the rhythms of teachers’ academic calendars. 

We also observed differences in how teachers articulated design feedback. One partner with prior training in design thinking provided specific, actionable suggestions, while another partner defaulted to broader descriptors such as ``user friendly'' or ``hard to use.'' Although all input was valuable, this variation suggests the opportunity for lightweight training to help codesign partners express more concrete, design-oriented feedback, without adding extra burden. \edit{Prior work similarly shows that even minimal, structured guidance can help people produce more specific and actionable design critiques \cite{Krause2017}.} Building reciprocal, sustainable, and effective codesign practices requires structuring collaborations that honor teachers’ time and rhythms, while also creating space for their growth as designers and learners.

\section{\edit{Conclusion}}
\edit{To support educational assessment authoring, we developed a conceptual model of teachers' workflows and a web-based system for authoring assessments through multilevel reusable interactions with \textsc{llm}s. Over seven months, we codesigned \ripplet with 13 teachers to (1) develop a conceptual model of their assessment authoring practices and derive a set of design objectives; (2) build a prototype and refine the interactions with teachers to reach the final version of \ripplet, which supports generating and reusing questions from diverse inputs, adapting questions at multiple granularities, restructuring assessments, as well as tracking and reapplying teachers' edits; and (3) create assessments for their students and observe their evolved usage and behavior over time. We found that \ripplet enabled teachers to create formative assessments they would not have otherwise made, shifted their practices from generation to curation, and encouraged their reflection on assessment quality. In a user study with 15 additional teachers, we compared \ripplet{} with their current practice. Teachers felt the results were worth the effort more and the assessment quality improved ($+1.93$ and $+1.11$ on a $10$-point scale, $p<0.05$). Together, we demonstrate \ripplet as a valuable educational tool that improves both teacher experience and assessment quality through its multilevel reusable interaction paradigm.\looseness=-10
}

\section*{Author Contributions}
Yuan Cui conceptualized and led the project, conducted all co-design sessions, designed and implemented the system, analyzed the data, and drafted the manuscript.
Annabel Marie Goldman contributed to system implementation, helped with user studies, and drafted part of the manuscript.
Jovy Zhou, Xiaolin Liu, Clarissa M. Shieh, Joshua Yao, and Mia Lillian Change contributed to system implementation.
Matthew Kay conceptualized the project, supervised the work, and edited the manuscript.
Fumeng Yang conceptualized and funded the project, supervised the work, contributed to system implementation, edited and drafted part of the manuscript.

\begin{acks}
This work would not have been possible without the generosity of our teacher partners in the codesign process and the user study. We are grateful for their time and invaluable feedback. We thank April Shi, Irena Liu, Laura Félix, Christopher Heo, Eric Lee, and Rachel Johnson for their early-stage contributions to building Ripplet. We extend our gratitude to Steven Moore, Mike Horn, Eleanor O’Rourke, and Duri Long for their feedback. 
\end{acks}
\bibliographystyle{ACM-Reference-Format}
\bibliography{ripplet}










\end{document}